\documentclass[aps,prd,onecolumn,preprintnumbers,nofootinbib,superscriptaddress]{revtex4}
\usepackage{graphicx}
\usepackage{bm,latexsym,amsmath,amssymb,amsfonts,color}

\begin{document}

\title{Viscoelastic and thermal response on nuclear pasta states at finite baryon density}

\author{Nicolás Grandi}
\email{grandi@fisica.unlp.edu.ar}
\affiliation{Departamento de Física "Dr. Emil Bose", Universidad Nacional de la Plata, Calle 49 y 115 s/n, CC67, 1900 La Plata, Argentina.}
\affiliation{Instituto de Física de La Plata, CONICET Diagonal 113 e/63 y 64, CC67, 1900 La Plata, Argentina.}

\author{Scarlett Rebolledo}
\email{rebolle3@msu.edu}
\affiliation{Department of Physics and Astronomy, Michigan State University, East Lansing, MI 48824, U.S.A.}

\author{Aldo Vera}
\email{aldo.vera@umayor.cl}
\affiliation{Núcleo de Matemática, Física y Estadística, Universidad Mayor, Avenida Manuel Montt 367, Santiago, Chile.}
\affiliation{Centro Multidisciplinario de Física, Vicerrectoría de Investigación, Universidad Mayor, Camino La Pirámide 5750, Santiago, Chile.}

\begin{abstract}
We compute the viscoelastic and thermal response of non-homogeneous hadronic condensates at finite baryon density representing crystals of baryonic tubes and layers. We describe them using analytic solutions of the Skyrme model in $3+1$ dimensions as ground states for a perturbative approach. At low enough temperatures, the lowest-energy excitations are described by a free massless scalar field theory in $1+1$ dimensions. We apply the Green-Kubo formulas to such excitations to obtain the elasticity tensor and other response coefficients. The analytic results of our computations are compared with available results on the nuclear pasta phase.
\end{abstract}

\maketitle

\newpage
\section{Introduction}
\label{sec:intro} 

One of the important open issues in theoretical physics is to achieve a satisfactory description of the phase diagram of Quantum Chromodynamics (QCD). The most challenging region of this phase diagram corresponds to low temperatures and finite baryon density. The common belief is that in this regime only refined numerical techniques can be effective (see \cite{newd3, newd4, newd5, newd6} and references therein), while analytical tools are often deemed insufficient. An unfortunate consequence of this preconception is that until very recently the appearance of the  \textit{nuclear pasta} phase \cite{pasta1, pasta2, pasta2a, pasta2b, pasta3, pasta4, pasta5, pasta6, pasta7, pasta8, pasta9, pasta10}, a remarkable phenomenon typical of such a regime, had no theoretical first-principles explanation. Moreover, the numerical analysis of these configurations is quite challenging   (see \cite{aprox0, aprox1, aprox2, aprox3, aprox4, aprox5, aprox6, aprox7, aprox8, aprox9, aprox10} and references therein) and requires  substantial computing power.

Crucial information on multi-baryonic configurations
is encoded in their elastic, viscous, and thermal response properties  \cite{pastacond1, pastacond2, pastacond3, pastacond4, pastacond5}. It is difficult to underestimate the relevance of these quantities, both in particle and nuclear physics as well as in astrophysics. As is clear from the above references, without a proper analytic understanding of the complex structures characterizing the nuclear pasta phase, the only way to compute these properties is through numerical simulations. Unfortunately,  extending the relevant numerical techniques to non-equilibrium regime is very difficult at the moment.  On the other hand, a ``theoretical dream" would be to have a proper analytic description of these multi-baryonic systems to which Green-Kubo formalism could be applied \cite{kubo1, kubo2, kubo3} (for a detailed pedagogical review see \cite{kubo4}). In this way, one would achieve a first principle understanding of these properties and would be able to make a 
comparison with the available experimental data.

The main goal of this paper is to take the first steps towards the theoretical dream mentioned above. Our starting point is the Skyrme theory \cite{skyrme}, which (at leading order in the 't Hooft expansion \cite{Gerard, largeN1, largeN2}) represents the low-energy limit of QCD. It is worth emphasizing that the Skyrme model is also compatible with Chiral Perturbation Theory, which is a more ``modern'' approach to the analysis of the low-energy limit of QCD  \cite{CHPT1, CHPT2, CHPT3}. Its action possesses both small excitations describing pions and topological solitons describing baryons \cite{Lizzi, shifman1, shifman2, witten0, ANW}, with the baryonic charge being a topological invariant (see also \cite{[8]p, [9]p, [10]p, [11]p, [12]p, sakurai, Machleidt, m1, m2, m3} and references therein).

In order to compute, via the Green-Kubo formalism, the viscoelastic and thermal response of regular multi-baryonic structures, we will use the analytic crystal-like solutions with high topological charge constructed in \cite{crystal0, crystal1, crystal2, crystal3, crystal4, crystal5, crystal6, crystal7, crystal8, crystal9} using the methods developed in \cite{grav1, grav2} (see also \cite{Cacciatori:2024mpf, Vera:2025qqz, euler4}). It is worth emphasizing that both the plots and the qualitative characteristics shown in \cite{crystal1}, \cite{crystal8} are very close to those found numerically in the analysis of ``spaghetti" configurations (see \cite{pasta1, pasta2, pasta2a, pasta2b, pasta3} and \cite{pasta10}). Moreover, in \cite{crystal5}, the shear modulus of ``lasagna-like" configurations has been computed, with the result being in good agreement with \cite{pasta5} and \cite{pasta9}. Thus, the present formalism is well-equipped to analyze the nuclear pasta phase.\footnote{Furthermore, the present results can be extended to the case of magnetized baryonic layers using the techniques introduced in \cite{charged1, charged2, Cacciatori:2025irb}. Quite interestingly, using recent results on the BPS equation in superfluids \cite{Canfora:2025jqr, Canfora:2025qkl} (see also \cite{Canfora:2024mkp}), it may be possible to apply the formalism proposed in the present manuscript to compute the transport coefficients in the Gross-Pitaevskii equation. We will return to both issues in a future publication.}

\section{The Skyrme model}
\label{sec:Skyrme.model}

The action for the $SU(2)$-Skyrme model can be written as
\begin{align}
    I[U] = \int 
d^{4}x    \sqrt{-g} \,
     \frac{K}{4}\,\text{Tr}\!\left[ R_{\mu }R^{\mu }+\frac{\lambda }{8}G_{\mu \nu }G^{\mu \nu }\right] \,.
\label{eq:action}
\end{align}
Here, the tensors $R_\mu$ and $G_{\mu\nu}$ are given, in terms of the Skyrme field $U(x)\in SU(2)$, as $R_{\mu }=U^{-1}\nabla_{\mu }U$ and $G_{\mu \nu }=[R_{\mu},R_{\nu }]$. The couplings $K$ and $\lambda $ are positive constants determined experimentally. 
The energy-momentum tensor arising from the above action takes the form
\begin{equation}
T_{\mu \nu} =-\frac{K}{2}\,\text{Tr}\!\left[ R_{\mu }R_{\nu }-\frac{1}{2}g_{\mu
\nu }R^{\alpha }\!R_{\alpha }+\frac{\lambda }{4}\!\left(  G_{\mu\alpha }G_{\nu}^{~ \alpha}\!-\!\frac{1}{4}g_{\mu \nu }G_{\sigma \rho }G^{\sigma \rho
} \!\right) \right] 
\,.
\label{eq:energy.momentum}
\end{equation}
There is also a topological charge $B$, which represents the baryon content of any given field configuration. It is defined on a spacelike hypersurface $\Sigma$, according to
\begin{equation}
B=\frac{1}{24\pi ^{2}}\int_{\Sigma }
\epsilon^{ijk}\,\text{Tr}\!\left[ R_{i}R_{j} R_{k} \right] 
\ .
\label{eq:topological.charge}
\end{equation}
We are interested in finite density effects; namely, we want to analyze the phenomena that occur when a non-vanishing baryonic charge is confined within a bounded spatial volume. Therefore, we consider the metric of a box as a starting point, whose line element is 
\begin{equation}
ds^{2}=-dt^{2}+L^{2}_x dx^{2}+ L^{2}_y dy ^{2}+ L^{2}_z dz ^{2} \ .  
\label{eq:metric}
\end{equation}
Here, the spatial Cartesian coordinates $x^i=(x,y,z)$ are dimensionless. and have the finite ranges $0\leq x\leq 2\pi$, $0\leq y \leq \pi$, $0\leq z \leq 2\pi$. The constants $L_x, L_y$ and $L_z$ represent the sides of the box where the solitons are confined, so that the total available volume is $4\pi ^{3}L_x L_y L_z$. We will work in the limit in which $L_x$ is much larger than the other two length scales $L_y, L_z$, even if the results presented in the present section and in section \ref{sec:spaghetti.and.lasagna} are valid for any value of these parameters. This separation of energy scales will be useful to simplify the perturbative treatment in Section \ref{sec:response}.

We will find it convenient in what follows to change variables in the plane $(t,x)$ to light-like coordinates $(x_+,x_-)$ defined according to
\begin{equation}
x_+=x+\frac{t}{L_x}
\qquad\mbox{and}\qquad
x_-=x-\frac{t}{L_x}\ .
\label{eq:lightlike.variables}
\end{equation}
In the forthcoming sections, we present exact solutions to the equations of motion derived from action \eqref{eq:action} and quantize the fluctuations around them. Then, we calculate the expectation values of the commutators of the different components of the energy-momentum tensor \eqref{eq:energy.momentum}. When inserted into the Green-Kubo formula, these commutators provide the elasticity and viscosity tensors and the thermal conductivities, as explained below.

\section{Inhomogeneous baryonic condensates: \emph{spaghetti} and \emph{lasagna} phases}
\label{sec:spaghetti.and.lasagna}

In this section, we briefly review the solutions representing tubular and layered multi-baryonic states. For a comprehensive discussion, the reader is referred to the original literature \cite{crystal0, crystal1, crystal2, crystal3, crystal4, crystal5, crystal6, crystal7, crystal8, crystal9}. As we shall see, even if we start with a different parametrization for each class of solutions, the resulting equations for the Ansatz functions are very similar.

\subsection{Crystals of baryonic tubes: the nuclear \emph{spaghetti} phase}
\label{sec:spaghetti} 

The standard exponential parametrization of the Skyrme field $U$  reads
\begin{equation}
U=\mathbf{1}_{2\times 2}\cos \alpha + n^{a}t_{a}\sin \alpha
\quad\quad
\mbox{with}\quad \ n^{a}=(\sin \Theta\, \cos \Phi ,\sin \Theta\, \sin \Phi , \cos \Theta) \ ,
\label{eq:spaghetti.parametrization}
\end{equation}
where $\mathbf{1}_{2\times 2}$ is the $2\times 2$ identity matrix, $t_i=i\sigma_i$ are the $SU(2)$ generators,  and $\alpha$, $\Theta$, $\Phi$ are the three spacetime-dependent scalar degrees of freedom of the Skyrme field $U$.

In terms of this parametrization, the topological charge \eqref{eq:topological.charge} reads  
\begin{equation}
B=\frac1{2\pi^2}\int dx\,dy\,dz\,\sin \Theta\,\sin^{2} \!\alpha \,\epsilon^{ijk}\, \partial_i\Phi\,
\partial_j\Theta\, \partial_k\alpha \ .
\label{eq:spaghetti.topological.charge}
\end{equation}
From this expression, it follows that to obtain non-trivial topological configurations, we must impose the necessary (but not sufficient) condition that $\alpha$, $\Theta$, and $\Phi$ are independent functions. 
Following \cite{crystal8}, let us consider the  Ansatz 
\begin{equation} 
\alpha =\alpha (z) \ ,   \qquad \Theta  =q \,y \ , \qquad \Phi =G(t,x) =G(x_+,x_-) \ , 
\label{eq:spaghetti.ansatz}
\end{equation}
where $q$ is an odd integer $q=2v+1$ with $v\in \mathbb{Z}$.  
Substituting this Ansatz into \eqref{eq:spaghetti.topological.charge}, the resulting topological charge is
\begin{eqnarray}
B&=& \frac 1{4\pi^2} \left.G(t,x)\right|_{x=0}^{x=2\pi} 
\left.\left(2\alpha -\sin(2\alpha)\right)\right|_{z=0}^{z=2\pi}
=np\, .
\label{eq:spaghetti.topological.charge.result}
\end{eqnarray}
Here, in the second equality, we have imposed the boundary conditions $\alpha (2\pi )=\alpha (0) +n\pi$ and $G(t,2\pi)=G(t,0)+2 \pi\, p$, with $n$ and $p$ being integer numbers.  Therefore, the topological charge is an arbitrary integer, implying that multi-baryonic solutions are well described by this Ansatz. 

Replacing the Ansatz into the Skyrme field equations obtained from the action in \eqref{eq:action}, they reduce to a single first order differential equation for the profile $\alpha(z)$, which can then be expressed as a first integral of the form
\begin{equation}
z=\frac{L_y}{q\,L_{z}} \left . \int \sqrt{ \frac{1+\gamma^2 \sin ^{2}\!\alpha}{e_n+\sin^2\!\alpha }}\ d\alpha \right .  \, .
\label{eq:spaghetti.alpha}
\end{equation}
Here $e_n$ is an integration constant fixed in terms of the integer $n$ by the boundary conditions, and we defined $\gamma^2={q^2\lambda}/L_y^2$. Furthermore, the equations of motion imply that the function $G(t,x)=G(x_+,x_-)$ must satisfy the constraint
\begin{equation} 
\left( \frac{\partial G}{\partial t}-\frac{1}{L_{x}}\frac{\partial G}{\partial x }\right)
\left( \frac{\partial G}{\partial t}+\frac{1}{L_{x}}\frac{\partial G}{\partial x}\right) =   \frac{\partial G}{\partial x_+} \frac{\partial G}{\partial x_-}=0\ .
\label{eq:spaghetti.G}
\end{equation}

This is solved by any function of a single light-like variable $G(x_+)$ or $G(x_-)$. It is important to highlight that this constraint, which plays a key role in the following discussion, arises from the inclusion of the Skyrme term and does not appear in the case of the NLSM  \cite{crystal8}.

When plugging the solution into the energy-momentum tensor \eqref{eq:energy.momentum}, an immediate conclusion can be drawn: The energy density $T_{tt}$ oscillates periodically in the $(y,z)$ directions. This allows us to interpret the solution as a crystal of parallel baryonic tubes extended along the $x$ axis, the so-called \emph{spaghetti} phase.

\subsection{Crystals of baryonic layers: the nuclear \emph{lasagna} phase}
\label{sec:lasagna}
As clarified in references \cite{crystal5} and \cite{crystal8}, baryonic layers are well described using the Euler angles parametrization for the $SU(2)$ field in the form
\begin{equation}
U=e^{t_{3}\Theta} \,e^{t_{2}\alpha}\,e^{t_{3}\Phi}
\ ,  
\label{eq:lasagna.parametrization}
\end{equation}
where now $\alpha$, $\Theta$ and $\Phi$ represent the three spacetime-dependent scalar degrees of freedom of the Skyrme field. Notice that, although we are using the same notation for the scalar functions as in the previous section, the parametrization itself is different.

The topological charge in the parametrization \eqref{eq:lasagna.parametrization} can be straightforwardly calculated and takes the form
\begin{equation}
B=\frac1{2\pi^2}\int dx\,dy\,dz\,\sin\left(\frac\alpha2\right) \,\epsilon^{ijk}   \,\partial_i\Phi \,\partial_j\Theta\, \partial_k\alpha \ .  
\label{eq:lasagna.topological.charge}
\end{equation}
Again, a non-vanishing topological charge requires independent functions. We choose the same functional dependencies for $\Theta$, $\alpha$, and $\Phi$ in the Ansatz as in the \textit{spaghetti} phase, namely those given in Eq. \eqref{eq:spaghetti.ansatz}.
Plugging the functions into the above expression for the topological charge \eqref{eq:lasagna.topological.charge}, we get
\begin{equation}
B=-\frac {q}{4\pi}\left.G\right|_{x=0}^{x=2\pi}\left. \cos (2\alpha)\right|_{z=0}^{z=2\pi} =qp\ ,
\end{equation}
where in the last equality we have imposed the boundary conditions for the scalar functions 
$\alpha(0)= 0$, $\alpha(2\pi) = (n+1/2) {\pi}$, and 
$G(t,0)=G(t,2\pi)+2 \pi\, p$ in terms of two  integer numbers $n,p\in \mathbb{Z}$. Recalling that $q$ is also an integer, the topological charge is an arbitrary integer, implying that the configuration describes a multi-baryonic state.

In the parametrization \eqref{eq:lasagna.parametrization}, the Ansatz simplifies the Skyrme field equations, reducing them to a single linear second order equation for the function $\alpha$, which is solved by
\begin{eqnarray}
\alpha(z)=\frac12\left( n+\frac12 \right) z  \ , 
\label{eq:lasagna.alpha}
\end{eqnarray}
where the integration constant has been determined by the boundary conditions. 

Once again, we obtain the constraint \eqref{eq:spaghetti.G} for the function $G(t,x)=G(x_+,x_-)$, which is then  solved by either  $G(x_+)$ or $G(x_-)$. 

The resulting energy density $T_{tt}$ is now periodic in the $z$ direction but independent of $y$. This allows for the interpretation of the solution as a crystal of parallel baryonic layers extended along the $(x,y)$ directions, which is called \emph{the lasagna} phase.  

\begin{figure}[t]
  \centering
\includegraphics[width=0.3\textwidth]{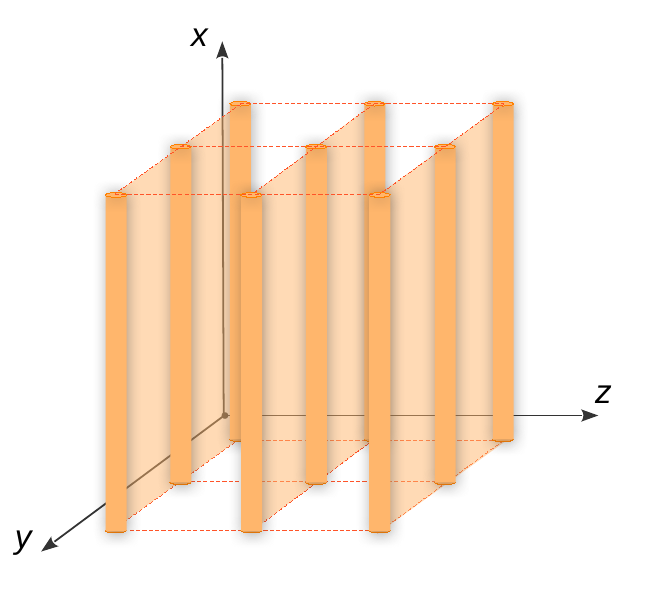}
  \caption{Schematic representation of the energy density profiles for the \emph{spaghetti} (cylinders) and \emph{lasagna} (planes) solutions. It is evident from the figure that the energy density is invariant in the $x$ direction, and for the \emph{lasagna} solution also in the $y$ direction. }
  \label{fig:your_label}
\end{figure}
\bigskip

A schematic representation of the solutions is shown in Fig. \ref{fig:your_label}. As should be clear from this short review, nuclear pasta states can be obtained quite simply from the Skyrme model once the appropriate parametrization is selected and a suitable Ansatz is found.

An important point in the forthcoming discussion is the identification  of the lowest energy  {\em spaghetti} and {\em lasagna} solutions. The inhomogeneous baryonic condensates are defined by the previously discussed Ansatz functions. The function $\Theta$ is described by the linear solution given in \eqref{eq:spaghetti.ansatz}, while function $\alpha$ depends solely on $z$ and takes the form shown in Eqs. \eqref{eq:spaghetti.alpha} and \eqref{eq:lasagna.alpha} for the \textit{spaghetti} and \textit{lasagna} backgrounds, respectively. Conversely, there is significant degeneracy in the choice of   the function $\Phi$, as any function $\Phi=G(x_-)$ depending only on $x_-$, or $\Phi=G(x_+)$ depending only on $x_+$, yields a classical solution. However, the energy density $T_{tt}$ obtained from \eqref{eq:energy.momentum} includes a contribution proportional to $(\partial_\pm \Phi)^2$. This suggests that the energy density is locally minimized by the linear expression $G(x_\pm)=px_\pm$. For definiteness, we will adopt the forward-propagating solution $G(x_-)=px_-$.

\section{Low energy excited states: emergent scalar field}
\label{sec:scalar.field}

Following linear response theory, we construct a quantum field theory describing the excitations around a nuclear pasta state by perturbing the classical solutions presented in the previous sections. 

The relevant question is now: what are the lowest-energy excitations?
In principle, various types of deformations are possible; we have three degrees of freedom ($\alpha$, $\Theta$, $\Phi$) that may depend on four space-time coordinates. 
On the one hand, in the limit where $L_x$ is much larger than the other two length scales $L_y$ and  $L_z$ (corresponding to a high density of {\em spaghetti}/{\em lasagna}), the typical energy of perturbations depending on $(y,z)$ (which is of order $1/L_y$ or $1/L_z$) is much higher than the typical energy of perturbations depending on $x$ (which is of order $1/L_x$). 
On the other hand, perturbations of the profiles $\alpha$ and $\Theta$ depending only on $x$ and time do not lead to consistent linearized equations, since the background solutions for $\alpha$ and $\Theta$ depend non-trivially on the other spatial coordinates.
The only perturbation that can consistently depend on $x$ is that of the profile $\Phi$.  
We then perturb the ground state with an off-shell perturbation of the form:
\begin{equation}
\Phi = p\,x_- + c \, \varphi(x_+,x_-) \ .
\label{eq:chiral.perturbations}
\end{equation}
Here, $c$ is a normalization constant to be determined below. It is important to note that the dependence of $\varphi(x_+,x_-)$ on $x_+$ can be made arbitrarily smooth, allowing the energy of the perturbations to be taken as close to the ground state energy as desired. Consequently, we do not perturb the functions $\Theta$ and $\alpha$, nor do we include any dependence on $y$ and $z$ in the perturbation $\varphi(x_+,x_-)$. This approach is justified by the fact that such perturbations would only contribute corrections at higher energy scales.

By inserting Eq. \eqref{eq:chiral.perturbations} into the Skyrme model action \eqref{eq:action} and expanding it to second order in the perturbation $\varphi(x_+,x_-)$, we obtain the dynamics for the excited states. This results in a scalar field action in $1+1$ dimensions for the variable $\varphi(x_+,x_-)$, multiplied by a factor arising from the integration over the $(y,z)$ coordinates:
\begin{equation}
I_{(2)} = \underbrace{ c^2 \int dy \, dz \, \Omega(y,z) }_{=1} \times \frac{1}{2} \int dt \, dx \, \left\{ (\partial_t \varphi)^2 - \frac{1}{L_x^2}(\partial_x \varphi)^2 \right\} \ . 
\label{eq:scalar.action.conformal}
\end{equation}
In this expression, the function $\Omega(y,z)$ characterizes the specific geometry of the {\em lasagna} or {\em spaghetti} solutions. We choose the constant $c$ in \eqref{eq:chiral.perturbations} such that the overall factor is unity, i.e., $c^{-2} = \int dy \, dz \, \Omega(y,z)$. This yields a canonical scalar action for the perturbation field $\varphi(x_+,x_-)$.

In conclusion, an emergent free $1+1$ dimensional scalar theory describes the low-energy perturbations of the ground state. This effective description significantly facilitates the calculation of the response functions, as detailed in the following section.

\section{Linear response of the \emph{nuclear pasta} states} 
\label{sec:response}
In this section, we aim to determine the linear response properties of the nuclear pasta states via several generalized susceptibilities, which can be straightforwardly calculated using the emergent scalar field described in the previous section. A brief discussion of the underlying theory, along with the relevant references, can be found in Appendix \ref{sec:appendix.response}.

As a first example, we are interested in the viscoelastic and thermoelastic response coefficients, which are defined by the following constitutive equation
\small
\begin{equation}
\sigma^{ij}(x)=\int d^2x'\left(E^{ijkl}(x,x')\, u_{kl}(x')+\eta^{ijkl}(x,x')\,\partial_t u_{kl}(x')
-\beta^{ij}(x,x')\,\delta T(x')-\tilde\theta^{ijk}(x,x')\partial_k\alpha(x')\right) \ .
\label{eq:viscoelastic}
\end{equation}
\normalsize
Here, $\sigma^{ij}$ represents the stress tensor, and $u_{ij}=(\partial_iu_j+\partial_j u_i)/2$ denotes the strain tensor, defined in terms of the deformation vector $u_i$. The stress and strain are related through the elasticity $E^{ijkl}$ and viscosity $\eta^{ijkl}$ tensors. We include the thermoelastic response under a temperature perturbation $\delta T$ via the thermal stress tensor $\beta^{ij}$. We also include an additional term that relates the stress tensor to the gradient of the \textit{thermal displacement} $\delta \alpha$, defined such that $\partial_t\delta\alpha= \delta T$. The variable $\delta \alpha$ is sometimes called \textit{thermasy}, and it can be interpreted as the mean quadratic dispersion of Brownian particles at time $t$ \cite{qlo}. The convolution integral is necessary only in the inhomogeneous directions $(y,z)$. Regarding the homogeneous directions $(x,t)$, they have been Fourier transformed to $(k,\omega)$, which has turned the convolution into a standard product. Formula \eqref{eq:viscoelastic} is valid in the small $\omega$ limit.

We are also interested in the thermal response, which we write in the form
\small
\begin{equation}
q^i(x)=\int d^2x'\left(-\kappa^{ij}(x,x')\,\partial_j\delta  T(x')+\iota^{ij}(x,x')\,\partial_j  \delta \alpha(x')+\pi^i(x,x')\delta T(x')+T(x')\theta^{ijk}(x,x')u_{jk}(x') \right)\, .
\end{equation}
\normalsize
Here, $\delta T$ represents the temperature perturbation, whose gradient contracts with the thermal conductivity tensor $\kappa^{ij}$ to yield the heat flux $q^i$. The additional terms correspond to an inductive component of the heat flow, which can be understood as a consequence of the inertia of convective currents \cite{thermal}. The response $\pi^i$ characterizes heat flow originating from a homogeneous change in temperature. This term is non-vanishing in a non-homogeneous material where the heat capacity varies from point to point. The final term takes into account the heat flow response induced by a strain. We show in Appendix \ref{sec:appendix.response} that the Maxwell-like relations $\tilde\theta^{ijk}(x,x')=-\theta^{ijk}(x',x)$ are satisfied.
 
Finally, we consider the response of the energy density under thermal deformations, given by
\begin{equation}
    \delta \rho(x')=\int d^2x'\left(c_V(x,x')\delta T(x')+\tilde \pi^i(x,x')\partial_i\delta\alpha(x') +\frac1T\tilde\beta^{ij}(x,x')u_{ij}(x') \right)\ ,
    \label{eq:thermal}
\end{equation}
where $c_V$ is the heat capacity, and we have included an additional term that accounts for the possibility of storing energy within a thermasy gradient. The last term represents the elastic energy. In Appendix \ref{sec:appendix.response}, we prove that $\tilde\pi^i(x,x')=-\pi^i(x',x)$ and $\tilde\beta^{ij}(x,x')=\beta^{ij}(x',x)$.

The response coefficients can be derived from a suitably defined spacetime tensor $\lambda_{\mu\nu\rho\sigma}$ in the low-frequency limit, using the following formulae (see Appendix \ref{sec:appendix.response} for details)
\begin{align}
\lambda^{ijkl} &= E^{jikl} + i \omega \,\eta^{ijkl} + \mathcal{O}(\omega^2)\,, & \lambda^{itjt} &= -T\left(\iota^{ij} + i\omega\, \kappa^{ij}\right) + \mathcal{O}(\omega^2) \,, \nonumber \\ 
\lambda^{ttti} &= -T\pi^i + \mathcal{O}(\omega) \,, & \lambda^{ittt} &= -T\tilde{\pi}^i + \mathcal{O}(\omega) \,, \nonumber \\
\lambda^{ijkt} &= T\theta^{ijk} + \mathcal{O}(\omega) \,, & \lambda^{tijk} &= T\tilde{\theta}^{ijk} + \mathcal{O}(\omega) \,, \nonumber \\
\lambda^{tttt} &= -T\,c_V + \mathcal{O}(\omega) \,. & &
\end{align}
The tensor  $\lambda_{\mu\nu\rho\sigma}$ can be obtained using the Kubo formula, which involves the commutators of the energy-momentum tensor components $T_{\mu\nu}$ as follows
\begin{equation} \label{eq:correlator.T}
{\lambda}_{\mu\nu\rho\sigma}(y,z,y',z';k,\omega) = -i \int_0^\infty \! dt \int_0^{2\pi} \! d x \, e^{i(\omega t - k x)} \left\langle \left[ T_{\mu\nu}(t,x,y,z), T_{\rho\sigma}(0,0,y',z') \right] \right\rangle \,. 
\end{equation}
To calculate the required integrals, we notice that in both the \textit{spaghetti} and \textit{lasagna} phases, the energy-momentum tensor in Eq. \eqref{eq:energy.momentum} takes the following generic form
\begin{align}
T_{zz} &= f_{zz}(y,z) + g_{zz}(y,z) \partial_+\Phi  \partial_-\Phi \,, & T_{xz} &= 0 \,, \, \nonumber \\
T_{yy} &= f_{yy}(y,z) + g_{yy}(y,z)  \partial_+\Phi  \partial_-\Phi \,, & T_{tz} &= 0 \,, \nonumber \\
T_{tx} &= g_{\cdot\cdot}(y,z) \left( (\partial_+\Phi)^2 - (\partial_-\Phi)^2 \right) \,, & T_{zy} &= 0 \,, \nonumber \\
T_{ty} &= g_{\cdot y}(y,z)(\partial_+\Phi - \partial_-\Phi) \,, & T_{xx} &= f_{\cdot\cdot}(y,z) + g_{\cdot\cdot}(y,z) \left( (\partial_+\Phi)^2 + (\partial_-\Phi)^2 \right) \,, \nonumber \\
T_{x y} &= g_{\cdot y}(y,z)  (\partial_+\Phi + \partial_-\Phi) \,, & T_{tt} &= -f_{\cdot\cdot}(y,z) + g_{\cdot\cdot}(y,z) \left( (\partial_+\Phi)^2 + (\partial_-\Phi)^2 \right) \,.
\label{eq:energy.momentum.generic}
\end{align}
Here, the \textit{form} functions $f_{zz}, f_{yy}, f_{\cdot\cdot}, g_{zz}, g_{yy}, g_{\cdot\cdot}$, and $g_{\cdot y}$ depend on whether the system is in the \textit{spaghetti} or \textit{lasagna} phase; their explicit forms are given in Section \ref{sec:form}.

With the help of Eq. \eqref{eq:energy.momentum.generic}, the different non-vanishing components of the tensor $\lambda_{\mu\nu\rho\sigma}$ can be obtained by quantizing the scalar field found in the previous section (see Appendix \ref{sec:appendix.scalar}) and then evaluating the necessary commutators (see Appendix \ref{sec:appendix.commutators}). 

We obtain the following non-vanishing elastic coefficients
\begin{align}
E_{xxxx} &= -g_{\cdot\cdot}g_{\cdot\cdot}' \left[ 2c^2 p^2 L_x^2 + \frac{c^4 L_x^3}{2\pi}\left( \frac{L_x}{2\pi}f_0(T) + W(k,T) \right) \right],
& E_{zzxx}  &= -\frac{c^2 L_x^4}{8\pi^2}\,f_0(T)\; g_{zz}g_{\cdot\cdot}', \nonumber \\[4pt]
E_{zzzz} &=- g_{zz}g_{zz}' \left[ \frac{c^2 p^2 L_x^2}{2} + \frac{c^4 L_x^3}{8\pi}\left( \frac{L_x}{2\pi}f_0(T) + w(k,T) \right) \right], 
& E_{yyxx}  &= -\frac{c^2 L_x^4}{8\pi^2}\,f_0(T)\; g_{yy}g_{\cdot\cdot}', \nonumber \\[4pt]
E_{yyyy} &=- g_{yy}g_{yy}' \left[ \frac{c^2 p^2 L_x^2}{2} + \frac{c^4 L_x^3}{8\pi}\left( \frac{L_x}{2\pi}f_0(T) + w(k,T) \right) \right], 
& E_{yyxy} &= -\frac{c^2 L_x^4}{8\pi^2}\,f_0(T)\; g_{yy}g_{\cdot y}', \nonumber \\[4pt]
E_{zzyy} &= -g_{zz}g_{yy}' \left[ \frac{c^2 p^2 L_x^2}{2} + \frac{c^4 L_x^3}{8\pi}\left( \frac{L_x}{2\pi}f_0(T) + w(k,T) \right) \right], 
& E_{xyxx}  &= -p c^2 L_x^2\; g_{\cdot y}g_{\cdot\cdot}',
\nonumber \\ 
E_{xyxy}  &= - c^2 L_x^2\; 
g_{\cdot y}g_{\cdot y}'\, ,&E_{xyzz} &= \frac{p c^2 L_x^2}{2}\; g_{\cdot y}g_{zz}'\,,
\end{align}
while the components of the viscosity tensor $\eta^{ijkl}$ vanish. Moreover, for the thermal stress tensor, we have
\begin{equation}
\begin{aligned}
\beta_{zz} &= -\frac{c^2 L_x^4}{8\pi^2}\,f_0(T)\; g_{zz}g_{\cdot\cdot}'\,, &\quad\quad \beta_{yy} &= -\frac{c^2 L_x^4}{8\pi^2}\,f_0(T)\; g_{yy}g_{\cdot\cdot}'\,, \\
\beta_{xy} &= -p c^2 L_x^2\; g_{\cdot y}g_{\cdot\cdot}'\,, & \beta_{xx} &= -g_{\cdot\cdot}g_{\cdot\cdot}' \left[ 2c^2 p^2 L_x^2 + \frac{c^4 L_x^3}{2\pi}\left( \frac{L_x}{2\pi}f_0(T) + W(k,T) \right) \right]\,.
\end{aligned}
\end{equation}
The thermal conductivity tensor $\kappa^{ij}$ vanishes, while for the thermal inductance, we obtain
\begin{align}
\iota_{xy} &= \frac{1}{T} p c^2 L_x^2\; g_{\cdot\cdot}g_{\cdot y}'\,, & \iota_{yy} &= \frac{1}{T} c^2 L_x^2\; g_{\cdot y}g_{\cdot y}'\,, \nonumber \\
\iota_{xx} &= \frac{1}{T} g_{\cdot\cdot}g_{\cdot\cdot}' \left[ 2c^2 p^2 L_x^2 + \frac{c^4 L_x^3}{2\pi}\left( \frac{L_x}{2\pi}f_0(T) + W(k,T) \right) \right]\,.
\end{align}
For the heat capacity, we get
\begin{equation}
c_V = \frac{1}{T}\,g_{\cdot\cdot}g_{\cdot\cdot}' \left[ 2c^2 p^2 L_x^2 + \frac{c^4 L_x^3}{2\pi}\left( \frac{L_x}{2\pi}f_0(T) + W(k,T) \right) \right]\,.
\end{equation}
The heat flux under a homogeneous temperature change is governed by the coefficients
\begin{align}
\pi^{x} &= -\frac{2}{T} p^2 c^2 L_x^2\; g_{\cdot\cdot}g_{\cdot\cdot}'\,, & \pi^{y} &= \frac{1}{T} p c^2 L_x^2\; g_{\cdot y}g_{\cdot\cdot}'\,, 
\end{align}
while the elastocaloric responses are
\begin{align} 
\theta^{x x x} &= \frac{2}{T} p^2 c^2 L_x^2\; g_{\cdot\cdot}g_{\cdot\cdot}'\,, & \theta^{x x y} &= -\frac{1}{T} p c^2 L_x^2\; g_{\cdot\cdot}g_{\cdot y}'\,, \nonumber \\[4pt]
\theta^{y x x} &= -\frac{1}{T} p c^2 L_x^2\; g_{\cdot y}g_{\cdot\cdot}'\,, & \theta^{z z y} &= \frac{1}{2T} p c^2 L_x^2\; g_{\cdot y}g_{zz}'\,.
\end{align}
Here, the prime on the form factors $g$ indicates that they are evaluated at $(z', y')$. These formulae contain the function
\begin{equation}
f_0(T) = \frac{1}{Z_0(T)} \sum_{n_0} e^{-\frac{n_0^2}{4\pi T}} n_0^2 \approx 
\begin{cases} 
0 & T \ll \frac{1}{4\pi} \\ 
2\pi T & T \gg \frac{1}{4\pi}
\end{cases} \, ,
\end{equation}
which is proportional to the free energy of a particle of mass $2\pi$ with periodic boundary conditions in a one-dimensional box of size $2\pi$, and corresponds to the zero mode of the emergent scalar field (see Appendix \ref{sec:appendix.scalar}). We also define the auxiliary functions
\begin{align}
W(k,T) &= \frac{1}{k} \sum_{k''} \coth\left(\frac{k''}{2L_x T}\right) k''(k-k'') \,, \\
w(k,T) &= \sum_{k''} \coth\left(\frac{k''}{2L_x T}\right) \frac{k''(k-k'')}{k - 2k''} \,,
\end{align}
which satisfy
\begin{equation}
W(k,T) \approx 2w(k,T) \approx 
\begin{cases} 
\mathsf{k}''^2_{\max} & T L_x \ll k \\ 
4 T L_x \mathsf{k}''_{\max} & 1 \ll k \ll T L_x
\end{cases} \, ,
\end{equation}
where $\mathsf{k}''_{\max}$ is a cutoff at which additional modes, not contained in the scalar field $\varphi$, are excited (see Appendix \ref{sec:appendix.commutators}). These would correspond to modes with momentum in the $(y,z)$ directions and/or oscillations of the $\alpha$ and $\Theta$ fields. As explained before, these modes are suppressed as the density of baryonic tubes/layers is large enough $L_x\gg L_y,L_z$, implying that we can make $\mathsf{k}''_{\max}$ as large as we want.  However, in the Skyrme model, there is a natural cutoff in the energy integrals, as it is an effective field theory describing the low-energy limit of QCD, valid up to energies of the order of 200 MeV. This last observation sets a physical interpretation for $\mathsf{k}''_{\max}$.

\section{Conformal factors and form functions}
\label{sec:form}

In this section, we calculate the normalization constant $c$ in \eqref{eq:chiral.perturbations} and the form functions $f_{zz}, f_{yy}, f_{\cdot\cdot}, g_{zz}, g_{yy}, g_{\cdot\cdot}$ and $g_{\cdot y}$ in \eqref{eq:energy.momentum.generic}. It is worth noting that these magnitudes have a purely classical origin and do not introduce any additional temperature dependence into the response functions.  These complete the results for the viscoelastic and thermal response of the nuclear pasta states discussed in the previous section.

\subsection{Response of the spaghetti phase}
\label{sec:response.spaghetti}

For the crystal of baryonic tube solutions described in Section \ref{sec:spaghetti}, inserting the Ansatz \eqref{eq:spaghetti.parametrization}-\eqref{eq:spaghetti.ansatz} into the Skyrme action \eqref{eq:action} yields the generic form \eqref{eq:scalar.action.conformal}. To obtain a standard action for a $1+1$ dimensional scalar, we impose a normalization constant $c$ in \eqref{eq:chiral.perturbations} with the value:
\begin{equation}
c^{-2} = \frac{\pi}{2L_x} \int_0^{2\pi} \! dz \, \sin^2 \alpha \left( K L_x L_y L_z + 4 \gamma^2 \frac{ e_n + \sin^2 \alpha (2 + \gamma^2 \sin^2 \alpha) }{ 1 + \gamma^2 \sin^2 \alpha } \right) \ .
\end{equation}
While this integral does not possess a closed analytic form, it can be evaluated numerically using the exact solution for $\alpha$ in \eqref{eq:spaghetti.alpha} for various parameter ranges.

On the other hand, the form functions in \eqref{eq:energy.momentum.generic} are obtained by substituting \eqref{eq:spaghetti.parametrization}-\eqref{eq:spaghetti.ansatz} into the expression for the energy-momentum tensor \eqref{eq:energy.momentum} and identifying the coefficients of the different powers of $\partial_\pm \Phi$. This results in

\small
\begin{align}
g_{yy} &= \frac{2KL_y^2}{L_x^2}\sin^2(qy)\sin^2 \alpha \frac{ \gamma^2(e_n + \sin^2 \alpha(1 - \gamma^2 \sin^2 \alpha) )-1}{ 1 + \gamma^2 \sin^2 \alpha } \,, \qquad\qquad\qquad\qquad\ g_{\cdot y} = 0 \,,
 \nonumber \\
g_{zz} &= \frac{2KL_z^2}{L_x^2} \sin^2(qy) \sin^2 \alpha \frac{ \gamma^2(e_n - \sin^2 \alpha(1 + \gamma^2 \sin^2 \alpha))-1 }{ 1 + \gamma^2 \sin^2 \alpha } \,, 
\ \nonumber 
\end{align}
\begin{align}
g_{\cdot\cdot} &= K \sin^2(qy) \sin^2 \alpha \frac{  \gamma^2(e_n + \sin^2 \alpha(3  + \gamma^2 \sin^2 \alpha)) +1}{ 1 + \gamma^2 \sin^2 \alpha } \,, &
\nonumber\\
f_{\cdot\cdot} &= -\frac{ K L_x^2 q^2 }{ 2 L_y^2 } (e_n + 2 \sin^2 \alpha) \,, \quad\quad
f_{zz} = \frac{K L_z^2q^2}{2L_y^2} e_n\, ,
\quad \quad f_{yy} = -\frac{Kq^2}{2} \frac{ e_n - \gamma^2 \sin^2 \alpha (e_n + 2 \sin^2 \alpha) }{ 1 + \gamma^2 \sin^2 \alpha } \,.
\label{eq:spaghetti.energy.momentum}
\end{align}
\normalsize

These form functions can now be used to obtain the specific viscoelastic and thermal response properties of the \textit{spaghetti} phase.

\subsection{Response of the lasagna phase}
\label{sec:response.lasagna}

For the crystal of baryonic layers discussed in Section \ref{sec:lasagna}, the field $\varphi(x_+,x_-)$ is canonically normalized in the action \eqref{eq:scalar.action.conformal} if the constant $c$ takes the value:
\begin{equation}
c^{-2} = \frac{4\pi^2}{L_x} \left( K L_x L_y L_z  + 2\gamma^2 + \frac{\lambda}{L_z^2} \left( n + \frac{1}{2} \right)^2 \right) \ . 
\end{equation}

The components of the energy-momentum tensor in lightlike coordinates yield the following form factors for the \textit{lasagna} phase:
\begin{small}
\begin{align}
g_{yy} &= \frac{2KL_y^2}{L_x^2} \left(  \gamma^2 \sin^2(2\alpha) - b_n^2 -1\right) \,, &
f_{yy} &= -\frac{KL_y^2}{2\lambda} \left( b_n^2 - (1 + b_n^2)\gamma^2 \right) \,, \nonumber \\
g_{zz} &= \frac{2KL_z^2}{L_x^2} \left( -\gamma^2 \sin^2(2\alpha) + b_n^2-1\right) \,, &
f_{zz} &= \frac{KL_z^2}{2\lambda} \left( b_n^2 - (1 - b_n^2)\gamma^2 \right) \,, \nonumber \\
g_{\cdot\cdot} &= K \left( 1 + b_n^2 + \gamma^2 \sin^2(2 \alpha) \right) \,, &
f_{\cdot\cdot} &= -\frac{KL_x^2}{2\lambda} \left( b_n^2 + \gamma^2(1 + b_n^2) \right) \,, \nonumber \\
g_{\cdot y} &= \frac{q K}{L_x} (1 + b_n^2) \cos(2\alpha) \,, &
\text{with} \quad b_n^2 &= \frac{\lambda}{4L_z^2} \left( n + \frac{1}{2} \right)^2 \,.
\end{align}
\end{small}

As in the previous case, neither the conformal factor nor the form functions introduce additional temperature dependence. The scaling argument also holds here: a larger box size maps into a smaller effective Skyrme coupling, placing the system in a perturbative regime.
\section{Discussion}
\label{sec:discussion}
In this paper, we have applied the Green-Kubo formalism to compute the viscoelastic, thermal, and thermoelastic responses of inhomogeneous baryonic condensates.
This  was made possible by the existence of analytic solutions representing hadronic crystals at finite baryon density within the Skyrme model. These solutions describe both crystals of baryonic tubes and layers and allow for the explicit computation of the Hamiltonian for the pionic excitations. This is particularly well-suited for the Kubo formalism, as it can be split into a static term and a time-dependent perturbation that can be turned on and off adiabatically. 

\begin{figure}[ht]
  \centering
\includegraphics[width=.68\textwidth]{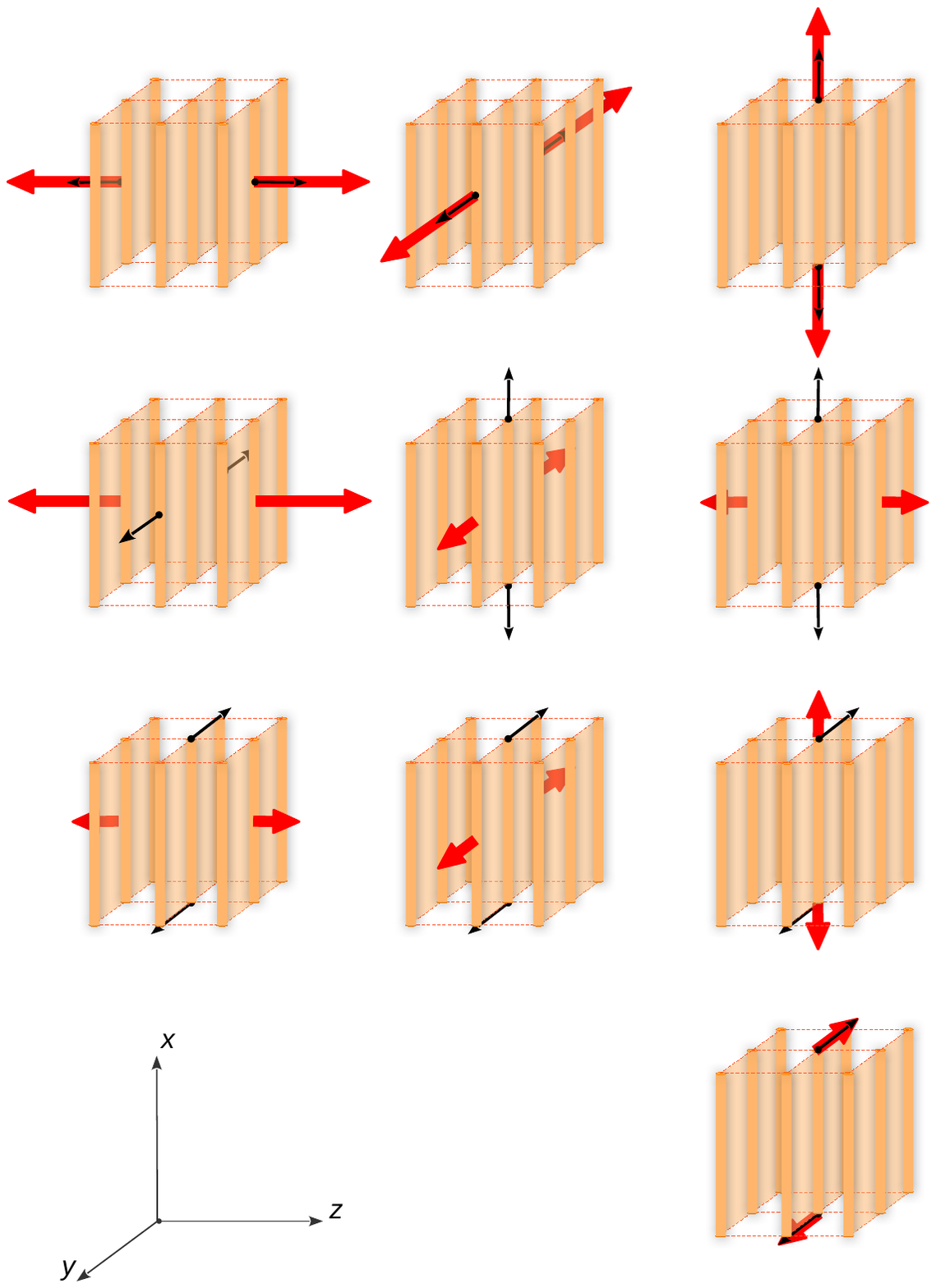}
\vspace{-0.9cm}
  \caption{Schematic representation elasticity tensor. Black arrows are the applied strain, and red arrows the resulting stress. 
In the second line we have the Poisson ratios, while
  the third and fourth lines represent shear ratios, all of them are small. }
  \label{fig:elastic.tensor}
\end{figure}

Our results for the elasticity tensor are summarized in the  schematic representation of Fig.~ \ref{fig:elastic.tensor}. Black arrows represent the applied strain, and red arrows represent the resulting stress. In the first line, we see the Young moduli, all of which are very large, as they grow with the cutoff as ${\sf k}''_{\sf max}$ or  ${{\sf k}''}^2_{\sf max}$ depending on the temperature regime.  In the second line, we have the Poisson ratios; the first one is very large, while the last two are small. The third and fourth lines represent shear ratios; all of them are small. 

A key point of this work is the topological origin of the rigidity of the nuclear pasta phases. The elasticity in our  approach is intrinsically linked to the baryonic charge distribution of the Skyrme model. The crystalline structure is topologically protected, meaning that the non-vanishing components of the elasticity tensor $E^{ijkl}$ arise from the fundamental symmetries and ``knots'' of the Skyrme field. This provides a first-principles justification for the Young moduli, the shear moduli, and the Poisson ratios of nuclear pasta, which is crucial for understanding the mechanical stability of the neutron star crust.

Furthermore, at the leading order of our effective field theory, the nuclear pasta states are described as an ideal elastic solid. While the viscosity $\eta$ and thermal conductivity $\kappa$ vanish in this free-field limit, the system remains a solid due to its resistance to shear deformations, distinct from a perfect fluid. The absence of dissipation at this level suggests that low-energy excitations propagate ballistically through the tubes and layers. 

Regarding the dissipative coefficients, although the free scalar field limit leads to vanishing viscosity and conductivity, we expect that higher-order corrections—such as self-interactions of the pionic excitations or couplings with other sectors like electrons and impurities—would lead to non-zero values. In such a regime, the transport coefficients are expected to exhibit a linear temperature scaling, $\eta, \kappa \sim T$, in agreement with several numerical results in the literature. 

Indeed, our analytical results present fundamental points of agreement with the existing macroscopic and geometric studies in the literature. On the one hand, the intrinsic stability of the non-homogeneous phases we describe agrees with the geometric analysis in \cite{Kubis}, which demonstrated that nuclear pasta phases are stable against shape perturbations and that the transition to other geometries requires crossing a finite energy barrier. On the other hand, regarding the dissipative transport properties, our findings support the molecular dynamics conclusions presented in \cite{pastacond2}, reaffirming that the non-spherical geometry of the pasta does not cause an anomalous increase in shear viscosity for purely morphological reasons, marking a clear distinction from the conventional behavior of terrestrial complex fluids. Thus, our results provide a rigorous baseline, representing the purely elastic limit of nuclear pasta and offering a clear path for future extensions of the model.

~

\subsection*{Acknowledgment}

The authors are grateful to Fabrizio Canfora for his collaboration and valuable contributions in the early stages of this work. S. R. has been funded by Fulbright-BIO 2021-56210068. A. V.
has been funded by FONDECYT Iniciaci\'on No. 11261883. N. G. is partially supported by CONICET grant PIP-2023-11220220100262CO, and
UNLP grant 2022-11/X931. 

\newpage
\appendix

\section{Geometry and linear response}
\label{sec:appendix.response}

Assume that a given physical system is perturbed at time $t=0$ by turning on a term in its Hamiltonian of the form
\begin{equation}
    \delta H = -J_a \mathcal{O}_a \, ,
\end{equation}
where $J_a$ is the \textit{source} of the operator $\mathcal{O}_a$. We define the \textit{response} of the perturbation on an operator $\mathcal{O}_b$ as the change in its expectation value $\delta\langle \mathcal{O}_b(\omega)\rangle$ induced by the perturbation. If the source is small enough, we can write to linear order
\begin{equation}
    \delta\langle \mathcal{O}_a(\omega)\rangle = -\alpha_{ab}(\omega) J_b(\omega) \, ,
\end{equation}
where $\alpha_{ab}(\omega)$ defines the \textit{generalized susceptibility}, whose components are often called the \textit{response coefficients}. Using time-dependent perturbation theory, we obtain its explicit form by means of the \textit{Kubo formula}
\begin{equation}
    \alpha_{ab}(\omega) = -i \int_0^\infty e^{i\omega t} \langle [\mathcal{O}_a(t), \mathcal{O}_b(0)] \rangle \, .
\end{equation}
In a system defined in a flat metric with coordinates $x^\mu$, a change $x^\mu \to x^\mu + \delta x^\mu$ can be absorbed into a metric deformation $\delta g_{\mu\nu} = \partial_\mu \delta x_\nu + \partial_\nu \delta x_\mu$. In the Hamiltonian, this implies
\begin{equation}
    \delta H = -\frac{1}{2} \int d^3x \, \delta g_{\mu\nu}(x) T^{\mu\nu}(x) \, ,
\end{equation}
where $T^{\mu\nu}$ is the stress-energy tensor. The response induced on the stress-energy tensor is then
\begin{equation}
    \delta \langle T_{\mu\nu}(\omega, x) \rangle = -\frac{1}{2} \int d^3x' \, \lambda_{\mu\nu}{}^{\rho\sigma}(\omega, x, x') \, \delta g_{\rho\sigma}(\omega, x') \, ,
    \label{eq:geometric.response}
\end{equation}
where the generalized susceptibility is given by
\begin{equation}
    \lambda_{\mu\nu\rho\sigma}(\omega, x, x') = -i \int_0^\infty dt \, e^{i\omega t} \langle [T_{\mu\nu}(t, x), T_{\rho\sigma}(0, x')] \rangle \, .
    \label{eq:geometric.Kubo}
\end{equation}

This formula is a powerful tool to obtain various response coefficients, including the elastic and viscosity tensors, thermal conductivity, heat capacity, and thermoelastic dissipation. 

The tensor $\lambda_{\mu\nu\rho\sigma}$ is symmetric under the exchanges $\mu \leftrightarrow \nu$ and $\rho \leftrightarrow \sigma$. Regarding its properties under the exchange $(\omega, x, \mu, \nu) \leftrightarrow (-\omega, x', \rho, \sigma)$, the tensor is {antisymmetric} when the total number of time-like indices ($0$) in the set $\{\mu, \nu, \rho, \sigma\}$ is {odd}, and {symmetric} when it is {even}. 

\subsection*{Viscoelastic response}

Under an elastic deformation, a material element at position $x^i$ is moved to $x^i+u^i(x,t)$, where $u^i(x,t)$ is the deformation vector. This deformation can be absorbed into a change of variables $x^i \to x^i-u^i(x,t)$. The resulting change in the metric is $\delta g_{ij}=-2 u_{ij}$, where $u_{ij}=(\partial_i u_j + \partial_j u_i)/2$ is the strain tensor. 

The corresponding response on the stress tensor $\sigma^{ij}=\langle \delta T^{ij}\rangle$ can be written using Eq. \eqref{eq:geometric.response} in the form
\begin{equation}
    \sigma^{ij}(\omega,x) = \int d^3x' \lambda^{ijkl}(\omega,x,x') \, u_{kl}(\omega,x') \ .
    \label{eq:response.strain}
\end{equation}
Assuming that $\lambda^{ijkl}(\omega,x,x')$ is finite for small frequencies, we can expand it as
\begin{equation}
    \lambda^{ijkl}(\omega,x,x') = E^{ijkl}(x,x') + i\omega\, \eta^{ijkl}(x,x') + \mathcal{O}(\omega^2) \ .
\end{equation}
Substituting this into \eqref{eq:response.strain}, it follows that the elasticity tensor $E^{ijkl}(x,x')$ characterizes the stress response to a static deformation, while the viscosity tensor $\eta^{ijkl}(x,x')$ corresponds to the response to the velocity of the deformation in the quasi-static limit. 

\subsection*{Thermal response}

A system at finite temperature $T$ is characterized by a planar Euclidean metric
\begin{equation}
    ds_E^2 = dt_E^2 + dx_i^2 \ ,
\end{equation}
where $t_E$ is the Euclidean time, which is periodic as $t_E \approx t_E + 1/T$. 

We consider a small temperature perturbation $T \to T + \delta T$, which modifies the Euclidean period to $1/(T + \delta T)$. By changing variables to a new Euclidean time $\tilde t_E = t_E (1 + \delta T/T)$, we recover the original period $\tilde t_E \approx \tilde t_E + 1/T$, but the metric is deformed to
\begin{equation}
    ds^2_E = \left(1 - \frac{2}{T}\delta T\right) d\tilde t_E^2 - \frac{2t_E}{T} \partial_i \delta T \, dx^i d\tilde t_E + dx_i^2 \ .
\end{equation}
Now we can Wick-rotate back to Minkowski spacetime to obtain a deformed spacetime metric
\begin{equation}
    ds^2 = \left(-1 + \frac{2}{T}\delta T\right) d\tilde t^2 + \frac{2t}{T} \partial_i \delta T \, dx^i d\tilde t + dx_i^2 \ .
\end{equation}
This results in a metric deformation with components $\delta g_{tt} = 2 \delta T/T$ and $\delta g_{ti} = t \partial_i \delta T/T$. Defining the thermal displacement (or \textit{thermasy}) $\alpha$ such that $\dot\alpha = T$, the latter can be rewritten as $\delta g_{ti} = \partial_i \delta \alpha/T$. 

\paragraph{Heat capacity:}
The above relations allow us to express the response of the energy density $\rho = T^{tt}$ using Eq. \eqref{eq:geometric.Kubo} as
\begin{equation}
    \delta \rho(\omega, x) = -\frac{1}{T} \int d^3x' \lambda^{tttt}(\omega,x,x') \, \delta T(\omega,x') - \frac{1}{T} \int d^3x' \lambda^{ttti}(\omega,x,x') \, \partial_i \delta \alpha(\omega,x') \ .
    \label{eq:delta.rho}
\end{equation}
The heat capacity at constant volume is identified from the static limit:
\begin{equation}
    \lambda^{tttt}(\omega,x,x') = -T \, c_V(x,x') + \mathcal{O}(\omega) \ .
    \label{eq:geometric.heat.capacity}
\end{equation}
Inserting this into \eqref{eq:delta.rho} yields a non-standard form due to the integral over $x'$. However, for an uniform temperature deformation ($\partial_i T = 0$), the standard heat capacity is recovered as the volume integral of $c_V(x,x')$ over $x'$.

We also codify the energy density response to a thermasy gradient via
\begin{equation}
    \lambda^{ttti}(\omega, x,x') = -T \, \tilde \pi^i(x,x') + \mathcal{O}(\omega) \ .
\end{equation}
The coefficient $\tilde\pi^i(x,x')$ represents an additional response, which is not frequently mentioned in the literature.

\paragraph{Thermal conductivity:}
The response equation for the energy current $q^i(x) = \delta T^{ti}(x)$ reads
\begin{equation}
    q^i(\omega,x) = -\frac{1}{T} \int d^3x' \lambda^{titj}(\omega,x,x') \, \partial_j \delta \alpha(\omega,x') - \frac{1}{T} \int d^3x' \lambda^{titt}(\omega,x,x') \, \delta T(\omega,x') \ .
    \label{eq:heat.transport}
\end{equation}
In the quasi-static limit, we define
\begin{equation}
    \lambda^{itjt}(\omega,x,x') = -T \left( \iota^{ij}(x,x') + i\omega \, \kappa^{ij}(x,x') \right) + \mathcal{O}(\omega^2) \ ,
\end{equation}
where $\kappa^{ij}(x,x')$ is the thermal conductivity. The term $\iota^{ij}(x,x')$ represents the \textit{thermal inductance}, characterizing the inertia of convective currents. 

An additional contribution proportional to $\delta T$ leads to the definition
\begin{equation}
    \lambda^{titt}(\omega, x,x') = -T \,  \pi^i(x,x') + \mathcal{O}(\omega) \ .
\end{equation}
This coefficient describes a heat flux induced by a homogeneous temperature change, arising from the material's spatial inhomogeneity. Notice that $\tilde \pi^i(x,x')=-\pi(x',x)$.

\subsection*{Thermoelastic response}

The stress induced by a temperature change is given by
\begin{equation}
    \sigma^{ij}(\omega,x) = -\frac{1}{T} \int \lambda^{ijtt}(\omega,x,x') \, \delta T(\omega,x') - \frac{1}{T} \int \lambda^{ijkt}(\omega,x,x') \, \partial_k \delta \alpha(\omega,x') \ .
\end{equation}
This allows us to define the thermal stress tensor $\beta^{ij}$ as
\begin{equation}
    \lambda^{ijtt}(\omega,x,x') = T \beta^{ij}(x,x') + \mathcal{O}(\omega) \ ,
\end{equation}
characterizing the stress at constant strain. The thermasy gradient contribution is captured by the terms
\begin{equation}
    \lambda^{ijkt}(\omega,x,x') = T \tilde\theta^{ijk}(x,x') + \mathcal{O}(\omega) \ .
\end{equation}

Conversely, the energy stored due to elastic deformation is studied via the energy density response:
\begin{equation}
    \delta \rho(\omega,x) = \int d^3x' \lambda^{ttij}(\omega,x,x') \, u_{ij}(\omega,x') \ ,
\end{equation}
with the expansion
\begin{equation}
    \lambda^{ttij}(\omega,x,x') = T \tilde\beta^{ij}(x,x') + \mathcal{O}(\omega) \ .
\end{equation}
Notice that here $\tilde\beta_{ij}(x,x')= \beta_{ji}(x',x)$.

Finally, the heat flux induced by strain is given by
\begin{equation}
    q^i = \int d^3x' \lambda^{tijk}(\omega,x,x') u_{jk}(\omega,x') \implies \lambda^{tijk}(\omega,x,x') = T \theta^{ijk}(x,x') + \mathcal{O}(\omega) \ .
\end{equation}
in terms of the {\em elastocaloric} response $\theta^{ijk}(x,x´)=-\tilde\theta^{ijk}(x',x)$.\footnote{Order $\omega$ terms were omitted here as they vanish in our specific nuclear pasta case, as shown in the following appendices.}

\section{Quantization of the emergent scalar field} 
\label{sec:appendix.scalar}

We assume that we have chosen $c$ (see equation \eqref{eq:chiral.perturbations}) in such a way that the scalar field $\varphi$ has action
\begin{equation}
    S=\frac12\int dt\,dx\,\left((\partial_t\varphi)^2-\frac1{L_x^2}(\partial_x\varphi)^2\right) \, .
\label{eq:scalar.action}
\end{equation}
The canonical momentum conjugated to $\varphi$ is then given by 
\begin{equation}
    \pi=\partial_t\varphi \, ,
\end{equation}
which allows us to write the canonical quantization rule
\begin{equation}
    [\varphi(t,x),\pi(t,x')]=i \sum_n\delta(x-x'-2n\pi) \, .
\label{eq:scalar.canonical}
\end{equation}
Here, the sum in the right hand side runs on $n\in \mathbb{Z}$, and it comes from the fact that we are quantizing the field in the cylinder $x\approx x+2\pi$, and thus we need a periodic Dirac $\delta$-function.

We decompose the fields in modes as
\begin{equation}
    \varphi=\sum_{k\neq0}\frac1{ 2\sqrt{\pi\omega_k}}\left(a_ke^{-i(\omega_kt-kx)}+a_k^\dagger e^{i(\omega_kt-kx)}\right)+\varphi_0(t) \, ,
\label{eq:scalar.decomposition}
\end{equation}
where $k\in\mathbb{Z}$, the dispersion relation is $\omega_k=|k|/L_x$, and $\varphi_0$ is the zero-mode. By plugging back into the action, we see that the zero-mode behaves as a free particle  with coordinate $\varphi_0$ and mass $2\pi$ (which comes from the $x$ integral). This allows us to write
\begin{equation}
    \varphi_0(t)=\frac{1}{2\pi}\,\Pi \,t+Q \, ,
\end{equation}
where $\Pi$ is the momentum conjugate to the coordinate $Q$ in the Schrödinger representation. This implies
\begin{equation}
    \pi= \sum_{k\neq0}\frac{-i\omega_k}{ 2\sqrt{\pi\omega_k}}\left(a_ke^{-i(\omega_kt-kx)}-a_k^\dagger e^{i(\omega_kt-kx)}\right)+\frac{1}{2\pi}\Pi \, .
\end{equation}
Inserting into the commutator and assuming $[\Pi,a_k]=[\Pi,a_k^\dagger]=[Q,a_k]=[Q,a_k^\dagger]=0$, as well as $[a_k,a^\dagger_{k'}]=\delta_{kk'}$, and $[Q,\Pi]=i$, we get
\begin{eqnarray}
[\varphi(t,x),\pi(t,x')]&=& 
\sum_{k\neq0} \frac{ i}{4\pi }
\left(
e^{-ik(x-x')}
+
e^{ik(x-x')}
\right) 
+\ \frac{i}{2\pi } \, .
\end{eqnarray}
In the second term inside the sum, we can change $k\to -k$ to have
\begin{equation}
    [\varphi(t,x),\pi(t,x')]=i \sum_{k\neq0}\frac{1}{2\pi}
e^{-ik(x-x')}+i\,\frac{1}{2\pi}=i \sum_{k}\frac{1}{2\pi}
e^{-ik(x-x')} \, ,
\end{equation}
where in the last equality we obtained the Fourier decomposition of the right-hand side of \eqref{eq:scalar.canonical}, showing that the normalization is consistent.

The Hamiltonian of the system reads
\begin{equation}
    H=\frac12\int dx\,\left((\partial_t\varphi)^2+\frac1{L_x^2}(\partial_x\varphi)^2\right)=\sum_{k\neq0}\omega_k \left(a^\dagger_k a_k+\frac12\right)+\frac{1}{4\pi}\Pi^2 \, .
\end{equation}

Operators $a^\dagger_k,a_k$ act as standard creation and annihilation operators in a vacuum state $|0\rangle$ of the $k$ modes, so $a_k^\dagger a_k=n_k$ is the occupation number, and they contribute to the energy as a collection of harmonic oscillators, as expected. Regarding the contribution from the zero mode, we quantize it as a free particle with wavefunction $\Psi(Q)$ on which the momentum acts as a derivative
\begin{equation}
    \Pi\Psi(Q)=-i\partial_Q \Psi(Q) \, .
\end{equation}
Then the energies are given by the Schrödinger equation
\begin{equation}
    -\frac{1}{4\pi}\partial_Q^2\Psi(Q)=E\Psi(Q) \, ,
\end{equation}
with solutions
\begin{equation}
    \Psi(Q)=\frac{1}{\sqrt{2\pi}}\,e^{i\sqrt{4\pi E}\,Q} \, ,
\end{equation}
where the normalization constant has been chosen so that the wavefunction is normalizable in the region $0<Q<2 \pi$. This is what matters since in both parametrizations \eqref{eq:spaghetti.parametrization} and \eqref{eq:lasagna.parametrization} our field is periodic in nature $\Phi\approx\Phi+2\pi$. This implies that $Q\approx Q+2\pi$, and then the energy gets quantized
\begin{equation}
    E_{n_0}= \frac{n_0^2}{4\pi} \,  ,
\end{equation}
where $n_0\in\mathbb{Z}$.
The eigenstates of the system then have energies given by
\begin{equation}
    E_{n_k}=\sum_{k\neq0}\left(n_k+\frac12\right)\omega_k+\frac{n_0^2}{4\pi} \, ,
\end{equation}
where also $n_k\in\mathbb{Z}$.
We can now use these energies to build the partition function as
\begin{eqnarray}
    Z&=&
    \left(\sum_{n_0}e^{-\frac{\beta n_0^2}{4\pi}}\right)
    \prod_k\left(\sum_{n_k}e^{-\beta\left(n_k+\frac12\right)\omega_k}\right)
    =Z_0(\beta)
    \prod_k Z_k(\beta) \, ,
\end{eqnarray}
where $Z_0(\beta)$ is the partition function of a particle in a box, and at low temperatures satisfies  $Z_0(\beta)\sim1+ e^{-\beta/4\pi}$, while at high temperatures is $Z_0(\beta)\sim 1/\sqrt\beta$. On the other hand, $Z_k(\beta)= 1/({2\sinh\left({\beta\omega_k}/2\right)})$ is the partition function of the oscillator modes. With this, we can write the expectation value of the number operator as
\begin{eqnarray}
    \langle a_k^\dagger a_k\rangle&=&
    \frac{1}{Z_k}\sum_{n_k} e^{-\beta\left( n_k+\frac12\right)\omega_k}n_k
    =
    - \frac1{\omega_k}\partial_\beta \log Z_k-\frac12
    \nonumber\\
    &=&\frac{1}{2}\left(
    \coth\left(\frac{\beta\omega_k}{2}\right)-1\right)=\frac1{e^{\beta\omega_k}-1}\equiv f_{BE}(\beta\omega_k) \, ,
\label{eq:scalar.occupation.mean}
\end{eqnarray}
as expected from Bose distribution. We also have, for the expectation value of the zero-mode energy, the following expression
\begin{equation}
    \langle \Pi^2\rangle=\frac1{Z_0(\beta)}\sum_{n_0}e^{-\frac{\beta n_0^2}{4\pi}}n_0^2=-4\pi \,\partial_\beta\log Z_0(\beta)
    =\left\{
    \begin{array}{cc}
         0& \mbox{low temperatures} \\
         &\\ 
         \frac{2\pi}{\beta} & \mbox{high temperatures}
    \end{array}
    \right\}\equiv f_0(\beta) \, .
\label{eq:scalar.energy.mean}
\end{equation}

With the results \eqref{eq:scalar.occupation.mean} and \eqref{eq:scalar.energy.mean}, we are in condition to calculate the propagator of the field. To do that, we write $\varphi=\varphi_++\varphi_-+\varphi_0$ where $\varphi_+$ contains only $a^\dagger_k$ and $\varphi_-$ contains only $a_k$, to get 
\begin{equation}
    \langle \varphi(t,x)\varphi(t',x')\rangle
    =
    \langle \varphi_+(t,x) \varphi_-(t',x')\rangle
    +
    \langle \varphi_-(t,x) \varphi_+(t',x')\rangle
    +
    \langle \varphi_0(t)\varphi_0(t')\rangle \, ,
\end{equation}
or, more explicitly
\begin{eqnarray}
    \langle \varphi(t,x)\varphi(t',x')\rangle
    &=&
    \sum_{k\neq0}\frac{1}{4\pi\omega_k}
    \left(
    \langle a_k^\dagger a_k\rangle
    e^{i(\omega_k(t-t')-k(x-x'))}+
    \langle a_k a_k^\dagger \rangle
    e^{-i(\omega_k(t-t')-k(x-x'))}
    \right)
    +
    \nonumber\\&&
    +\,
    \frac{1}{(2\pi)^2}\langle \Pi^2\rangle\,t\,t'
    +
    \frac1{2\pi}\left(\langle \Pi Q\rangle \, t+\langle Q\Pi\rangle t'\right)
    +
    \langle Q^2\rangle \, .
\end{eqnarray}
We can write the above in terms of the expectation values \eqref{eq:scalar.occupation.mean} and \eqref{eq:scalar.energy.mean}, in the form
\small
\begin{eqnarray}
    \langle \varphi(t,x)\varphi(t',x')\rangle
    &=&
    \sum_{k\neq0}\frac{1}{4\pi\omega_k}
    \left(
    e^{i(k(x-x')-\omega_k(t-t'))}
    { +}\,
    2f_{BE}(\beta\omega_k)\,
    {\cos}(k(x-x')-\omega_k(t-t'))
    \right)
    +
    \nonumber\\&&
    +\,
    \frac{1}{(2\pi)^2}f_0(\beta)\,t\,t'
    +
    \frac1{2\pi}\left(\langle \Pi Q\rangle \, t+\langle Q\Pi\rangle t'\right)
    +
    \langle Q^2\rangle \, ,
\end{eqnarray}
\normalsize
which can be rewritten in light-like coordinates as 
\small
\begin{eqnarray}
    \langle \varphi(t,x)\varphi(t',x')\rangle
    &=&
    \sum_{k>0}\frac{L_x}{4\pi k}
    \left(
    e^{i k\Delta x_-}+e^{-i k\Delta x_+}
    +
    2f_{BE}(\beta k/L_x)
    \left(
    \cos(k\Delta x_-)+\cos(k\Delta x_+)
    \right)
    \right)
    +
    \nonumber\\&&
    +\,
    \frac{1}{(2\pi)^2}f_0(\beta)\,t\,t'
    +
    \frac1{2\pi}\left(\langle \Pi Q\rangle \, t+\langle Q\Pi\rangle t'\right)
    +
    \langle Q^2\rangle \, .
\end{eqnarray}
\normalsize
We can use a  shorthand notation for this result, writing
\begin{eqnarray}
    \langle \varphi(t,x)\varphi(t',x')\rangle
    &=&
    \sum_{s=+,-}
    \sum_{k>0}\frac{L_x}{4\pi k}
    \left(
    e^{-is k\Delta x_s}
    +
    2 f_{BE}(\beta  k /L_x)\,
    \cos(k\Delta x_s)
    \right)+
        \nonumber\\&&
    +\,
    \frac{L_x^2}{(4\pi)^2}f_0(\beta)\,(x_+-x_-)(x'_+-x'_-)+
    \nonumber\\&&
    +
    \frac{L_x}{4\pi}\left(\langle \Pi Q\rangle \, (x_+-x_-)+\langle Q\Pi\rangle (x_+'-x_-')\right)
    +
    \langle Q^2\rangle \, .
\end{eqnarray}
Notice that no physical quantity could depend on the terms in the third line, since the expectation values $\langle Q^2\rangle$, $\langle Q\Pi\rangle$ and $\langle \Pi Q\rangle$ are not well defined on a circle. Since our observables depend on second derivatives of the above correlator, those constant and linear terms disappear and we can omit them in what follows. 

The above expression can be used to calculate the derivatives we need,  namely
\small
\begin{eqnarray}
    \langle \partial_{s_1}\varphi(t,x)\partial_{s_2}\varphi(t',x')\rangle
    &=&
    \delta_{s_1s_2}
    \sum_{k>0}\frac{k{L_x}}{4\pi}
    \left(
    e^{-is_1 k\Delta x_{s_1}}
    +
    2 f_{BE}(\beta  k /L_x)\,
    \cos(k\Delta x_{s_1})
    \right) 
    +\,
    \frac{L_x^2}{(4\pi)^2}f_0(\beta)\,s_1s_2 \, ,\nonumber\\
\end{eqnarray}
\normalsize
allowing us to write
\begin{eqnarray}
    \langle \partial_{s_1}\varphi(t,x)\partial_{s_2}\varphi(t',x')\rangle
    &=&
    \delta_{s_1s_2}K_{s_1}^{\sf odd}(\Delta x_{s_1})+K_{s_1s_2}^{\sf even}(\Delta x_{s_1}) \, ,
\end{eqnarray}
where we have
\begin{align}
&K_{s}^{\sf odd}=
-{ \frac{i {s}
L_{x}}{4\pi}} \sum_{k>0}{k} 
\sin(k\Delta x_s)
=
-{ \frac{ {s}
L_{x}}{8 \pi}} \ 
\sum_{k }k\, 
e^{i k\Delta x_s}
=
{ \frac{ i 
{s}
L_{x}}{4}}\, \sum_n\delta'({ s}\Delta x_s-2n\pi) \, ,
\\
&K_{s_1s_2}^{\sf even}=
{}
\frac{ L_{x}}{8 \pi}\delta_{s_1s_2}
  \sum_k
k\coth\left(\frac{\beta k}{2L_x}\right)e^{ik\Delta x_{s_1}}+    \frac{L_x^2}{({ 4}\pi)^2}f_0(\beta)\,s_1s_2 \, .
\label{eq:KoddKeven}
\end{align} 
In other words
\begin{equation}
    \langle [\partial_{s_1}\varphi(t,x),\partial_{s_2}\varphi(t',x')]\rangle
    =2\delta_{s_1s_2}K_{s_1}^{\sf odd}  \, .
    \label{eq:scalar.commutator.linear}
\end{equation}

We can now go further by evaluating
\begin{equation}
    \langle[ \partial_{s_1}\varphi\partial_{s_2}\varphi\,,\,\partial_{s_3}\varphi'\partial_{s_4}\varphi']\rangle 
    =
    \substack{\lim\\ 1\to 2}\,\substack{\lim\\3\to4}\,
\partial^1_{s_1}\partial^2_{s_2}\partial_{s_3}^3\partial_{s_4}^4
\,\langle
[\varphi_1\varphi_2\,,\,\varphi_3\varphi_4]
\rangle \, ,
\end{equation}
where we made a point-splitting to simplify the calculation. Using Wick theorem, we can rewrite
\begin{eqnarray}
\langle
\varphi_1\varphi_2\varphi_3\varphi_4
\rangle
&=& 
\langle\varphi_1\varphi_2\rangle
\langle\varphi_3\varphi_4\rangle
+
\langle\varphi_1\varphi_3\rangle
\langle\varphi_2\varphi_4\rangle
+
\langle\varphi_1\varphi_4\rangle
\langle\varphi_2\varphi_3\rangle \, .
\end{eqnarray}
Applying the derivatives and antisymmetrizing, we get
\begin{eqnarray}
\!\!\!\!\!\!\!\!\!\!\!\!\!\!\!\!\!
\langle
[\partial_{s_1}\varphi_1\partial_{s_2}\varphi_2\,,\,\partial_{s_3}\varphi_3\partial_{s_4}\varphi_4]
\rangle
&=&  
\langle\partial_{s_1}\varphi_1\partial_{s_3}\varphi_3\rangle
\langle\partial_{s_2}\varphi_2\partial_{s_4}\varphi_4\rangle
\!-\!
\langle\partial_{s_3}\varphi_3\partial_{s_1}\varphi_1\rangle
\langle\partial_{s_4}\varphi_4\partial_{s_2}\varphi_2\rangle
\nonumber\\&&
+ 
\langle\partial_{s_1}\varphi_1\partial_{s_4}\varphi_4\rangle
\langle\partial_{s_2}\varphi_2\partial_{s_3}\varphi_3\rangle
\!-\!
\langle\partial_{s_3}\varphi_3\partial_{s_2}\varphi_2\rangle
\langle\partial_{s_4}\varphi_4\partial_{s_1}\varphi_1\rangle \ .
\end{eqnarray}
Now, taking the coincidence limit independently on each entry of the commutator, we get
\begin{eqnarray}
\!\!\!\!\!\!\!\!\!\!
\langle
[\partial_{s_1}\varphi\,\partial_{s_2}\varphi\,,\,\partial_{s_3}\varphi'\partial_{s_4}\varphi']
\rangle
&=&  
\langle\partial_{s_1}\varphi\,\partial_{s_3}\varphi'\rangle
\langle\partial_{s_2}\varphi\,\partial_{s_4}\varphi'\rangle
-
\langle\partial_{s_3}\varphi'\partial_{s_1}\varphi\rangle
\langle\partial_{s_4}\varphi'\partial_{s_2}\varphi\rangle
\nonumber\\&&\,
+\,
\langle\partial_{s_1}\varphi\,\partial_{s_4}\varphi'\rangle
\langle\partial_{s_2}\varphi\,\partial_{s_3}\varphi'\rangle
-
\langle\partial_{s_3}\varphi'\partial_{s_2}\varphi\rangle
\langle\partial_{s_4}\varphi'\partial_{s_1}\varphi\rangle \, ,
\end{eqnarray}
so, we end up with
\small
\begin{eqnarray}
\langle
[\partial_{s_1}\varphi\,\partial_{s_2}\varphi\,,\,\partial_{s_3}\varphi'\partial_{s_4}\varphi']
\rangle
&=&  
2\left(
\delta_{s_2s_4}K_{s_1s_3}^{\sf even}K_{s_2}^{\sf odd}
+
\delta_{s_1s_3}K_{s_2s_4}^{\sf even}K_{s_1}^{\sf odd}
+
\delta_{s_2s_3}K_{s_1s_4}^{\sf even}K_{s_2}^{\sf odd}
+
\delta_{s_1s_4}K_{s_2s_3}^{\sf even}K_{s_1}^{\sf odd}
\right) \, .
\nonumber\\
\label{eq:scalar.commutator.quadratic}
\end{eqnarray}
\normalsize
This equation will be useful in the next section when evaluating the commutators of the components of the energy-momentum tensor.
\section{Commutators of the energy-momentum tensor}
\label{sec:appendix.commutators}
For our Anzätze \eqref{eq:spaghetti.ansatz} for \emph{spaghetti} \eqref{eq:spaghetti.parametrization} and \emph{lasagna} \eqref{eq:lasagna.parametrization} phases, the energy-momentum tensor takes the generic form
\begin{align}
&\!\!\!
T_{zz}=  f_{zz}(y,z)
+g_{zz}(y,z)\,\partial_+\Phi\,\partial_-\Phi  \, ,
& & \!\!
T_{x z}=0\,, 
\phantom{\frac{{1} }{2}}\, 
\nonumber\\
&\!\!\!
T_{yy}=  f_{yy}(y,z)
+g_{yy}(z,y)\,\partial_+\Phi\,\partial_-\Phi  \, ,
  & & \!\!
T_{tz}= 0\,,
\phantom{\frac{{1} }{2}}
\nonumber
\\
&\!\!\!
T_{tx}= g_{\cdot\,\cdot}(z,y)\left((\partial_+\Phi)^2-(\partial_-\Phi)^2\right)\,, \phantom{\frac{{1} }{2}}
& &  \!\!
T_{zy}=0 \, ,
\nonumber\\
&\!\!\! 
T_{ty}= 
g_{\cdot\,y}(y,z)\,(\partial_+\Phi-\partial_-\Phi)\,,
& & \!\!
T_{xx}= 
f_{\cdot\,\cdot}(y,z)+g_{\cdot\,\cdot}(y,z)\left((\partial_+\Phi)^2+(\partial_-\Phi)^2\right)\,,
\phantom{\frac{{1} }{2}}
\nonumber
\\
&\!\!\!
T_{x y}=
g_{\cdot\,y}(y,z)\,(\partial_+\Phi+\partial_-\Phi)\,, 
& & \!\!
T_{tt}= -f_{\cdot\,\cdot}(y,z)+g_{\cdot\,\cdot}(y,z)\left((\partial_+\Phi)^2+(\partial_-\Phi)^2\right)\,, 
\phantom{\frac{{1} }{2}}  \label{eq:energy.momentum.generic.Appendix}
\end{align}
where the form functions $f_{zz}, f_{yy}$, $f_{\cdot\,\cdot}$, $g_{zz}$, $g_{yy}$, $g_{\cdot\,\cdot}$ and $g_{\cdot\,y}$ depend on the phase, being different for the \emph{lasagna} and \emph{spaghetti} cases; see Section \ref{sec:form}. The  expectation value of the commutators of these energy-momentum components will be given in terms of the  expressions
\begin{align}
&\left\langle\left[ \partial_{+}\Phi\pm \partial_{-}\Phi, \partial_{+}  {\Phi^{\prime}} \pm \partial_{-}  {\Phi^{\prime}} \right] \right\rangle \, ,
\nonumber\\&
\left\langle\left[ (\partial_{-} \Phi)^2 + (\partial_{+} \Phi)^2, \partial_{+} \Phi^{\prime} \partial_{-}  \Phi^{\prime} \right]\right\rangle \, ,
\nonumber\\
&\left\langle\left[ \partial_{+}\Phi \pm \partial_{-}\Phi , (\partial_{-} \Phi')^2 \pm' (\partial_{+} \Phi')^2\right] \right\rangle \, ,
\nonumber\\
&\left\langle\left[ \partial_{+}\Phi \pm \partial_{-}\Phi , \partial_{-} \Phi' \,\partial_{+} \Phi ' \right] \right\rangle \, ,
\nonumber\\
&\left\langle\left[ \partial_{-} \Phi\, \partial_{+} \Phi , \partial_{-}  \,\Phi^{\prime} \partial_{+}  \Phi^{\prime} \right]\right\rangle \, ,
\nonumber\\&
\left\langle\left[ (\partial_{-} \Phi)^2 \pm (\partial_{+} \Phi)^2, (\partial_{-}  {\Phi^{\prime}})^2 \pm' (\partial_{+}  {\Phi^{\prime}})^2  \right]\right\rangle\!  \, ,
    \label{eq:commutators.relevant}
\end{align}
which can be evaluated by using the expressions \eqref{eq:scalar.commutator.linear} and \eqref{eq:scalar.commutator.quadratic} in terms of the functions \eqref{eq:KoddKeven}. To do that, we use the expression \eqref{eq:chiral.perturbations} for the field $\Phi$ in terms of the emergent scalar $\varphi$;
\begin{eqnarray}
\Phi&=&p\,x_-+ c \, \varphi(x_+,x_-)\ .
\end{eqnarray}
Inserting into \eqref{eq:commutators.relevant}, this allows us to calculate the relevant commutators needed to compute response functions, namely
\begin{align}
\left\langle\left[ \partial_{+}\Phi \pm \partial_{-}\Phi , \partial_{+}  {\Phi^{\prime}} \pm' \partial_{-}  {\Phi^{\prime}} \right] \right\rangle
&=
\left\langle\left[ c\,\partial_{+}\varphi \pm p\pm c\,\partial_{-}\varphi , c\,\partial_{+}  {\varphi^{\prime}} \pm' p\pm' c\,\partial_{-}  {\varphi^{\prime}} \right] \right\rangle
\nonumber\\
&=c^2\left\langle\left[ \partial_{+}\varphi \pm  \partial_{-}\varphi , \partial_{+}  {\varphi^{\prime}} \pm'  \partial_{-}  {\varphi^{\prime}} \right] \right\rangle
\nonumber\\
&=c^2\left\langle\left[ 
\partial_{+}\varphi ,\partial_{+}  \varphi^{\prime}\right]\right\rangle 
\pm(\pm')c^2\left\langle\left[\partial_{-}  {\varphi}.\partial_{-}  {\varphi^{\prime}} \right] \right\rangle
\nonumber\\
&=2c^2\left(K^{\sf odd}_{+}\pm(\pm')K^{\sf odd}_{-}\right) \, ,
\end{align}

\begin{align}
\left\langle\left[ (\partial_{-} \Phi)^2 \pm (\partial_{+} \Phi)^2, \partial_{+} \Phi^{\prime} \partial_{-}  \Phi^{\prime} \right]\right\rangle
&= 
\left\langle\left[ (p+c\,\partial_{-} \varphi)^2 \pm c^2\partial_{+} \varphi^2, c\,\partial_{+} \varphi^{\prime} (p+c\,\partial_{-}  \varphi^{\prime} \right]\right\rangle  
\nonumber\\
&   
\!\!\!\!\!\!\!\!\!\!\!=
\left\langle\left[ p^2+2pc\,\partial_{-} \varphi+c^2\partial_{-} \varphi^2 \pm c^2\partial_{+} \varphi^2,pc\,\partial_+\varphi'+c^2\partial_+\varphi'\partial_-\varphi' \right]\right\rangle  
\nonumber\\
& 
\!\!\!\!\!\!\!\!\!\!\!=
c^2\left(2p^2
\left\langle\left[ 
\partial_{-} \varphi,
\partial_+\varphi'
\right]\right\rangle
+
c^2\left\langle\left[
\partial_{-} \varphi^2 \pm \partial_{+} \varphi^2,\partial_+\varphi'\partial_-\varphi' \right]\right\rangle\right)  
\nonumber\\
& 
\!\!\!\!\!\!\!\!\!\!\!=
c^4\left\langle\left[
\partial_{-} \varphi^2 \pm  \partial_{+} \varphi^2,\partial_+\varphi'\partial_-\varphi' \right]\right\rangle
\nonumber\\
& 
\!\!\!\!\!\!\!\!\!\!\!=
4c^4\left(K_{-+}^{\sf even}K_{-}^{\sf odd}
\pm
K_{+-}^{\sf even}K_{+}^{\sf odd}
\right) \, ,
\end{align}
%
\begin{align}
\left\langle\left[ \partial_{+}\Phi \pm \partial_{-}\Phi , (\partial_{-} \Phi')^2 \pm' (\partial_{+} \Phi')^2\right] \right\rangle
&=
\left\langle\left[ c\,\partial_{+}\varphi \pm p\pm c\,\partial_{-}\varphi , (p+c\,\partial_{-}  {\varphi^{\prime}})^2 \pm' c^2\partial_{+}  {\varphi^{\prime}}^2    \right] \right\rangle
\nonumber\\
&=
\left\langle\left[ c\,\partial_{+}\varphi \pm p\pm c\,\partial_{-}\varphi , p^2+2pc\,\partial_{-}  {\varphi^{\prime}}  \right] \right\rangle
\nonumber\\
&=
\left\langle\left[ c\,\partial_{+}\varphi  \pm c\,\partial_{-}\varphi ,  2pc\,\partial_{-}  {\varphi^{\prime}}  \right] \right\rangle
\nonumber\\
&=\pm 4pc^2 K_{-}^{\sf odd} \, ,
\end{align}
\begin{align}
\left\langle\left[ \partial_{+}\Phi \pm \partial_{-}\Phi , \partial_{-} \Phi' \,\partial_{+} \Phi ' \right] \right\rangle
&=
\left\langle\left[ c\,\partial_{+}\varphi \pm p\pm c\,\partial_{-}\varphi , c\, (p+c\,\partial_-\varphi')\,c\,\partial_+\varphi'   \right] \right\rangle
\nonumber\\
&=\left\langle\left[ c\,\partial_{+}\varphi  \pm c\,\partial_{-}\varphi , (p+c\,\partial_-\varphi')\,c\,\partial_+\varphi'   \right] \right\rangle
\nonumber\\
&=pc^2\left\langle\left[ \partial_{+}\varphi  \pm \partial_{-}\varphi , \,\partial_+\varphi'   \right] \right\rangle
\nonumber\\
&=2pc^2 
    K_{+}^{\sf odd}  \, ,
\end{align}
\begin{align}
\left\langle\left[ \partial_{-} \Phi\, \partial_{+} \Phi , \partial_{-}  \,\Phi^{\prime} \partial_{+}  \Phi^{\prime} \right]\right\rangle
&=\left\langle\left[(p+c\,\partial_-\varphi)\,c\,\partial_+\varphi,(p+c\,\partial_-\varphi')\,c\,\partial_+\varphi'\right] \right\rangle=  \nonumber\\
&=\left\langle\left[pc\,\partial_+\varphi+c^2\,\partial_-\varphi\, \partial_+\varphi,pc\,\partial_+\varphi'+c^2\,\partial_-\varphi'\, \partial_+\varphi'\right] \right\rangle  =\nonumber\\&=
p^2c^2\,\left\langle\left[\partial_+\varphi, \partial_+\varphi'\right]\right\rangle+c^4\,\left\langle\left[\partial_-\varphi\,\partial_+\varphi,\partial_-\varphi'\, \partial_+\varphi'\right]\right\rangle
    , \nonumber\\
&=2c^2\left(p^2K^{\sf odd}_{+}
+
c^2
K_{--}^{\sf even}K_{+}^{\sf odd}
+
c^2
K_{++}^{\sf even}K_{-}^{\sf odd}
\right) \, ,
\end{align}
\small
\begin{align} 
\left\langle\left[ (\partial_{-} \Phi)^2 \pm (\partial_{+} \Phi)^2, (\partial_{-}  {\Phi^{\prime}})^2 \pm' (\partial_{+}  {\Phi^{\prime}})^2  \right]\right\rangle 
&=\left\langle\left[ (p+c\,\partial_{-} \varphi)^2 \pm c^2\partial_{+} \varphi^2, (p+c\,\partial_{-}  {\varphi^{\prime}})^2 \pm' c^2\partial_{+}  {\varphi^{\prime}}^2  \right]\right\rangle
\nonumber\\
&
\!\!\!\!\!\!\!\!\!\!\!\!
\!\!\!\!\!\!\!\!\!\!\!\!
\!\!\!\!\!\!\!\!\!\!\!\!
\!\!\!\!\!\!\!\!\!\!\!\!
\!\!\!\!\!\!\!\!
=
\left\langle\left[ p^2+2pc\,\partial_{-} \varphi+c^2\partial_{-} \varphi^2 \pm c^2\partial_{+} \varphi^2, 
p^2+2pc\,\partial_{-}  {\varphi^{\prime}}+c^2\partial_{-}  {\varphi^{\prime}}^2 \pm' c^2\partial_{+}  {\varphi^{\prime}}^2  \right]\right\rangle
\nonumber\\
&
\!\!\!\!\!\!\!\!\!\!\!\!
\!\!\!\!\!\!\!\!\!\!\!\!
\!\!\!\!\!\!\!\!\!\!\!\!
\!\!\!\!\!\!\!\!\!\!\!\!
\!\!\!\!\!\!\!\!
=
c^2\left(
4p^2\left\langle\left[ \partial_{-} \varphi,
\partial_{-}  {\varphi^{\prime}}
\right]\right\rangle
+c^2\left\langle\left[
\partial_{-} \varphi^2 \pm \partial_{+} \varphi^2, 
\partial_{-}  {\varphi^{\prime}}^2 \pm' \partial_{+}  {\varphi^{\prime}}^2  \right]\right\rangle
\right)
\nonumber\\
&
\!\!\!\!\!\!\!\!\!\!\!\!
\!\!\!\!\!\!\!\!\!\!\!\!
\!\!\!\!\!\!\!\!\!\!\!\!
\!\!\!\!\!\!\!\!\!\!\!\!
\!\!\!\!\!\!\!\!
=
c^2\left(
4p^2\left\langle\left[ \partial_{-} \varphi,
\partial_{-}  {\varphi^{\prime}}
\right]\right\rangle
+c^2
\left(
\left\langle\left[
\partial_{-} \varphi^2 ,
\partial_{-} \varphi'^2 
\right]\right\rangle
\pm'
\left\langle\left[
\partial_{-} \varphi^2 ,
\partial_{+} \varphi'^2
\right]\right\rangle
\pm
\right.\right.
\nonumber\\
&
\qquad\qquad
\left.\left.
\pm
\left\langle\left[
\partial_{+} \varphi^2,
\partial_{-} \varphi'^2 
\right]\right\rangle
\pm(\pm')
\left\langle\left[
\partial_{+} \varphi^2 ,
\partial_{+} \varphi'^2 
\right]\right\rangle
\right)\right)
\nonumber\\
&
\!\!\!\!\!\!\!\!\!\!\!\!
\!\!\!\!\!\!\!\!\!\!\!\!
\!\!\!\!\!\!\!\!\!\!\!\!
\!\!\!\!\!\!\!\!\!\!\!\!
\!\!\!\!\!\!\!\!
=
c^2\left(
4p^2\left\langle\left[ \partial_{-} \varphi,
\partial_{-}  {\varphi^{\prime}}
\right]\right\rangle
+c^2
\left(
\left\langle\left[
\partial_{-} \varphi^2 ,
\partial_{-} \varphi'^2 
\right]\right\rangle   
\pm(\pm')
\left\langle\left[
\partial_{+} \varphi^2 ,
\partial_{+} \varphi'^2 
\right]\right\rangle
\right)\right)
\nonumber\\
&
\!\!\!\!\!\!\!\!\!\!\!\!
\!\!\!\!\!\!\!\!\!\!\!\!
\!\!\!\!\!\!\!\!\!\!\!\!
\!\!\!\!\!\!\!\!\!\!\!\!
\!\!\!\!\!\!\!\!
=
c^2\left(
8p^2K_{--}^{\sf odd}
+4c^2
\left(
K^{\sf even}_{--}K^{\sf odd}_{-}
\pm(\pm')
K^{\sf even}_{++}K^{\sf odd}_{+}
\right)\right) \, .
\end{align}
\normalsize
In summary, we have 
\small
\begin{align}
&\left\langle\left[ \partial_{+}\Phi\pm \partial_{-}\Phi, \partial_{+}  {\Phi^{\prime}} \pm' \partial_{-}  {\Phi^{\prime}} \right] \right\rangle
=2c^2\left(K^{\sf odd}_{+}\pm(\pm')K^{\sf odd}_{-}\right) \, ,
\nonumber\\&
\left\langle\left[ (\partial_{-} \Phi)^2 \pm (\partial_{+} \Phi)^2, \partial_{+} \Phi^{\prime} \partial_{-}  \Phi^{\prime} \right]\right\rangle
=
4c^4\left(K_{-+}^{\sf even}K_{-}^{\sf odd}
\pm
K_{+-}^{\sf even}K_{+}^{\sf odd}
\right) \, ,
\nonumber\\
&\left\langle\left[ \partial_{+}\Phi \pm \partial_{-}\Phi , (\partial_{-} \Phi')^2 \pm' (\partial_{+} \Phi')^2\right] \right\rangle
=\pm 4pc^2K_{-}^{\sf odd} \, ,
\nonumber\\
&\left\langle\left[ \partial_{+}\Phi \pm \partial_{-}\Phi , \partial_{-} \Phi' \,\partial_{+} \Phi ' \right] \right\rangle
= 2pc^2 
    K_{+}^{\sf odd} \, ,
\nonumber\\
&\left\langle\left[ \partial_{-} \Phi\, \partial_{+} \Phi , \partial_{-}  \,\Phi^{\prime} \partial_{+}  \Phi^{\prime} \right]\right\rangle
=2c^2\left(p^2K^{\sf odd}_{+}
+
c^2
K_{--}^{\sf even}K_{+}^{\sf odd}
+
c^2
K_{++}^{\sf even}K_{-}^{\sf odd}
\right) \, ,
\nonumber\\&
\left\langle\left[ (\partial_{-} \Phi)^2 \pm (\partial_{+} \Phi)^2, (\partial_{-}  {\Phi^{\prime}})^2 \pm' (\partial_{+}  {\Phi^{\prime}})^2  \right]\right\rangle\! =\!
c^2\!\left(
8p^2K_{-}^{\sf odd}
+{ 8}c^2\!
\left(
K^{\sf even}_{--}K^{\sf odd}_{-}
\pm\!(\pm')
K^{\sf even}_{++}K^{\sf odd}_{+}
\right)\!\right) \, .
\label{eq:conmutadores.Appendix}
\end{align}
\normalsize
To obtain the generalized susceptibility \eqref{eq:correlator.T}, we need the Fourier transform of the commutators of the energy-momentum tensor. Since the commutators are given in terms of the expressions \eqref{eq:conmutadores.Appendix}, this implies that we need to compute the following relevant integrals:
\begin{align}
  &R_1^{\pm\pm'}=-i\int_0^{2\pi}\!\!dx\,\int_0^\infty \! dt \
  e^{i(\omega t-kx)}\left\langle\left[ \partial_{+}\Phi \pm \partial_{-}\Phi , \partial_{+}  {\Phi^{\prime}} \pm '\partial_{-}  {\Phi^{\prime}} \right] \right\rangle \, ,
    \nonumber\\
&R_2^{\pm}=-i\int_0^{2\pi}\!\!dx\,\int_0^\infty \! dt \ 
  e^{i(\omega t-kx)}\left\langle\left[  (\partial_{-} \Phi)^2 \pm (\partial_{+} \Phi)^2, \partial_{-}  \,\Phi^{\prime} \partial_{+}  \Phi^{\prime} \right]\right\rangle \, ,
      \nonumber\\
&R_3^{\pm\pm'}=-i\int_0^{2\pi}\!\!dx\,\int_0^\infty \! dt \
  e^{i(\omega t-kx)} \left\langle\left[ \partial_{+}\Phi \pm \partial_{-}\Phi , (\partial_{-} \Phi')^2 \pm' (\partial_{+} \Phi')^2\right] \right\rangle \, ,
    \nonumber\\
&R_4^{\pm}=-i\int_0^{2\pi}\!\!dx\,\int_0^\infty \! dt\
  e^{i(\omega t-kx)} \left\langle\left[ \partial_{+}\Phi \pm \partial_{-}\Phi , \partial_{-} \Phi' \,\partial_{+} \Phi ' \right] \right\rangle \, ,
    \nonumber\\
&R_5^{\pm\pm'}=-i\int_0^{2\pi}\!\!dx\,\int_0^\infty \! dt\ 
  e^{i(\omega t-kx)}\left\langle\left[ (\partial_{-} \Phi)^2 \pm (\partial_{+} \Phi)^2, (\partial_{-}  {\Phi^{\prime}})^2 \pm' (\partial_{+}  {\Phi^{\prime}})^2  \right]\right\rangle \, ,
  \nonumber\\
    &R_6=-i\int_0^{2\pi}\!\!dx\,\int_0^\infty \! dt\  e^{i(\omega t-kx)}\left\langle\left[ \partial_{-} \Phi\, \partial_{+} \Phi , \partial_{-}  \,\Phi^{\prime} \partial_{+}  \Phi^{\prime} \right]\right\rangle \, .
    \label{eq:integrals.relevant}
\end{align}
As can be checked by inspection of \eqref{eq:conmutadores.Appendix}, these integrals are composed of those elementary ones
\begin{align}
&E_1^s=-i\int_0^{2\pi}\!\!dx\,\int_0^\infty \! dt\ 
  e^{i(\omega  t-kx)}\,K_s^{\sf odd}\,,
\nonumber\\
&
E_2^{ss_1s_2}=  -i\int_0^{2\pi}\!\!dx\,\int_0^\infty \! dt\    e^{i(\omega t-kx)}\,K_s^{\sf odd}
K_{s_1s_2}^{\sf even} \, .
\end{align}
Consequently,
\begin{align}
    &R_1^{\pm\pm' }=2c^2(E_1^+\pm(\pm')E_1^-) \, ,
      \nonumber\\
        &R_2^{\pm}=4{ c^4
        }\left(E_2^{--+}\pm E_2^{++-}\right) \, ,
    \nonumber\\
&R_3^{+\pm'}=-R_3^{-\pm'}\equiv R_3 =4pc^2E_1^- \, ,
\nonumber\\
    &R_4^{\pm}\equiv R_4=2pc^2E_1^+
     \, ,\nonumber\\
    &R_5^{\pm\pm'}=2c^2\left(
    { 4}p^ 2E_1^-+4c^2\left(E_2^{---}\pm(\pm')E_2^{+++}\right) \right)\, ,
\nonumber\\
    &R_6=2c^2 \left(p^2E_1^++c^2\left(E_2^{+--}+E_2^{-++}\right)\right) \, .
    \label{eq:integrals.relevant.elementary}
\end{align}
The first elementary integral $E_1^s$ can be calculated as
\small
\begin{align} 
E_1^s&=-i\int_0^{2\pi}\!\!dx\,\int_0^\infty \! dt\ 
  e^{i(\omega t-kx)}\,K_s^{\sf odd}
  \nonumber\\
&=
i{ \frac{{ s}
L_{\phi}}{8 \pi}} \int_0^{2\pi}\!\!dx\,\int_0^\infty \! dt\      e^{i(\omega t-kx)}\,
 \sum_{k' }k'\, 
e^{i k'\left(x+\frac{s}{L_x}t\right)}
\nonumber\\ 
&= 
i { \frac{{ s}
L_{x}}{8 \pi}} \ \sum_{k' }\,k' \int_0^{2\pi}\!\!dx\ e^{i(k'-k)x}\int_0^\infty \! dt\     
e^{i \left(\omega+\frac{k's}{L_x}\right)t}
\nonumber\\ 
&= 
i { \frac{{ s}
L_{x}}{8 \pi}} \ \sum_{k' }\,k' \int_0^{2\pi}\!\!dx\ e^{i(k'-k)x}\left(
\frac{i}{\omega+\frac{k's}{L_x}}+\pi \,
\delta 
\left(\omega+\frac{k's}{L_x}\right)\right)
\nonumber\\ 
&= 
i { \frac{{ s}
L_{x}}{8 \pi}} \ \sum_{k' }\,k' \int_0^{2\pi}\!\!dx\ e^{i(k'-k)x}
\frac{i}{\omega+\frac{k's}{L_x}+i\,\varepsilon}
\nonumber\\ 
&= 
{ \frac{{ -s}
L_{x}}{4}}  
\frac{k}{\omega+\frac{ks}{L_x}+i\,\varepsilon}\, , \\
E_1^s&\approx
{-}\frac{ { 
}
{ L_{x}^2}}{4} { \equiv E_1} \, .
\end{align}
\normalsize
Regarding the second elementary integral $E_2$,  we get
\small
\begin{align}
E_2^{ss_1s_2}&=      -i\int_0^{2\pi}\!\!dx\,\int_0^\infty \! dt\    e^{i(\omega t-kx)}\,K_s^{\sf odd}
K_{s_1s_2}^{\sf even}
\nonumber\\
&=      -i\int_0^{2\pi}\!\!dx\,\int_0^\infty \! dt\    e^{i(\omega t-kx)}
\left(- { \frac{ {  s}
L_{x}}{8 \pi}}   \sum_{k' }k'\, 
e^{i k'\left(x+\frac{st}{L_x}\right)}\right) \nonumber\\
&~~~\times\left(   \frac{L_x^2}{({ 4}\pi)^2}f_0(\beta)\,s_1s_2 { {\, +\,} L_{x}} \delta_{s_1s_2}
  \sum_{ k}
 \frac{k}{8\pi}\coth\left(\frac{\beta k}{2L_x}\right)e^{ik\left(x+\frac{s_1t}{L_x}\right)}\right) \, .
\end{align}
\normalsize 
This integral has two terms corresponding to each of the terms in the parenthesis in the second line. The first term is immediately
\small
\begin{align}
        {}
        i { \frac{{ s}
        L_{x}}{8  \pi}}
        {\sum_k k'}
        \int_0^{2\pi}\!\!dx\,\int_0^\infty \! dt\    e^{i((\omega{+k's/L_x} )t-(k{-k'})x)}\,\frac{L_x^2}{({ 4}\pi)^2}f_0(\beta)\,s_1s_2=&  - { {s}
        }\frac{L_x^{ 3 
        } {k}}{4({  4}\pi)^2}f_0(\beta)
        \frac {s_1s_2}{{\omega+k\frac s{L_x}+i\varepsilon}}
        \nonumber\\
        &\approx 
        {
        -\frac{L_x^4}{(8\pi)^2}f_0(\beta)s_1s_2 \, ,
        }
\end{align}
\normalsize
and for the second term
\small
\begin{align}&
{ 
}i{ \frac{{s}
L_{x}^2}{ (8 \pi)
{ ^2}}}\delta_{s_1s_2}\int_0^{2\pi}\!\!dx\,\int_0^\infty \! dt\    e^{i(\omega t-kx)}
  \sum_{k' }k'\, 
e^{i k'\left(x+\frac{st}{L_x}\right)} 
  \sum_{ k''}
 {k''}\coth\left(\frac{\beta k''}{2L_x}\right)e^{ik''\left(x+\frac{s_1t}{L_x}\right)}
\nonumber\\
 &=
 { { }}
 i{ \frac{ { s}
 L_{x}^2}{(8 \pi)
{ ^2}}}\delta_{s_1s_2}\int_0^{2\pi}\!\!d x\, 
  \sum_{k' k''}k'\, 
e^{i (k'+k''-k)x} 
 k''\coth\left(\frac{\beta k''}{2L_x}\right) \int_0^\infty \! dt\   e^{i\left(\omega+k'\frac{s}{L_x} +k''\frac{s_1}{L_x}\right)t} 
 \nonumber\\
 &=
 {{ } 
 }i{ 
 \frac{ { s}
 L_{x}^2}{(8 \pi)
{ ^2}}}\delta_{s_1s_2} 
  \sum_{k' k''}k'\, 
 {k''}\coth\left(\frac{\beta k''}{2L_x}\right) 
 \frac { i}{\omega+k'\frac{s}{L_x} +k''\frac{s_1}{L_x}+i\varepsilon} \int_0^{2\pi}\!\!dx\, e^{i (k'+k''-k)x} 
  \nonumber\\
 &=
 { { -}
 }
 { \frac{(2 \pi){ s}
 L_{x}^2}{(8 \pi)
{ ^2}}}\delta_{s_1s_2}
  \sum_{k''}(k-k'')\, 
 {k''}\coth\left(\frac{\beta k''}{2L_x}\right) 
 \frac 1{\omega+(k-k'')\frac{s}{L_x} +k''\frac{s_1}{L_x}+i\varepsilon}  
   \nonumber\\
 &\approx
{ { -}
}
{ \frac{ 
 L_{x}^3}{ { 8}(2 \pi)
 }} \delta_{s_1s_2}
  \sum_{k''}\, 
 \coth\left(\frac{\beta k''}{2L_x}\right) 
 \frac {k''(k-k'')}{k -k''(1-ss_1) }  \equiv { { -} 
 }
 { \frac{ 
 L_{x}^3}{{ 32\pi}
 }}\delta_{s_1s_2} w_{ss_1}(k,\beta) \, .
\end{align}
\normalsize 
Thus, the second elementary integral reads
\small
\begin{align}
      E_2^{ss_1s_2}= - \frac{
      L_x^3}{32 \pi
      }
\left(
 { \frac{L_x}{2\pi }}f_0(\beta)s_1s_2 
+\delta_{s_1s_2}\,w_{ss_1}(k,\beta)
 \right) \, .
\end{align}
\normalsize 
In terms of the divergent sum
\begin{align}
w_{ss_1}(k,\beta)=  
  \sum_{k''}\, 
 \coth\left(\frac{\beta k''}{2L_x}\right) 
 \frac {k''(k-k'')}{k -k''(1-ss_1) }  \, .
\end{align}
The divergence arises at large values of $\kappa''$, which correspond to short wavelength fluctuations. In this limit, modes of the original Skyrme theory, which are not contained in the emergent scalar field $\varphi$, would show up. Thus, we have to cut-off the above sum at $|k''|<k''_{\sf max}$.

For $k\neq0$, the sum can be rewritten as
\begin{align}
w_{ss_1}(k,\beta)=    {  2}
  \sum_{k''\geq 0}^{k''_{\sf max}}\,
\coth\left(\frac{\beta k''}{2L_x}\right) k''
\frac { k^2-k''^2(1-ss_1) }{k^2 -k''^2(1-ss_1)^2}    \, .
\end{align}

Defining $\kappa=k''/k$, we can write
\begin{align}
w_{ss_1}(k,\beta)=    {  2 k}
  \sum_{\kappa\geq 0}^{\kappa_{\sf max}}\,
\coth\left(\frac{k\beta \kappa}{2L_x}\right) \kappa
\frac { 1-\kappa^2(1-ss_1) }{1 -\kappa^2(1-ss_1)^2}    \, .
\end{align}
If $k\gg 1$, the sum can be approximated by the integral
\begin{align}
w_{ss_1}(k,\beta)=   {  2 k^2}
  \int_0^{\kappa_{\sf max}}d\kappa\,
\coth\left(\frac{k\beta \kappa}{2L_x}\right) \kappa
\frac { 1-\kappa^2(1-ss_1) }{1 -\kappa^2(1-ss_1)^2}    \, .
\end{align}
We do not have an analytic expression for this integral, but we can approximate it in different limits. In the region $\kappa\gg 2L_x/\beta k$, we have $\coth(k\beta\kappa/2L_x)\approx 1$, while for $\kappa\ll 2L_x/\beta k$ we get $\coth(k\beta\kappa/2L_x)\approx 2L_x /k \beta \kappa$. This allows us to rewrite the integral as a sum, in the form
\begin{align}
w_{ss_1}(k,\beta)\approx   {  2 k^2}
\left(
\frac{2L_x }{\beta k} 
  \int_0^{\frac{2L_x}{\beta k}}d\kappa\,
\frac { 1-\kappa^2(1-ss_1) }{1 -\kappa^2(1-ss_1)^2}    
+ 
\int_{\frac{2L_x}{\beta k}}^{\kappa_{\sf max}}d\kappa\,
\kappa
\frac { 1-\kappa^2(1-ss_1) }{1 -\kappa^2(1-ss_1)^2}  
\right) \, .
\end{align}
If $k\gg TL_x$, only the second term remains
\begin{align}
w_{ss_1}(k,\beta)\approx    {  2 k^2}
  \int_0^{\kappa_{\sf max}}d\kappa\,
\kappa
\frac { 1-\kappa^2(1-ss_1) }{1 -\kappa^2(1-ss_1)^2}     \, ,
\end{align}
resulting in
\begin{align}
    &w_{++}(k,\beta)=w_{--}(k,\beta)={ }k''^2_{\sf max} \ ,
    \nonumber\\
    &w_{+-}(k,\beta)=w_{-+}(k,\beta)={  \frac{1}{2}}\left( k''^2_{\sf max}
    -\frac{i\pi }{ 4}k^2 \right) \, .
\end{align}
On the other hand, if $1\ll k\ll TL_x$, we get  
\begin{align}
w_{ss_1}(k,\beta)\approx    {  \frac{4 L
_x k}{\beta }}
  \int_0^{\kappa_{\sf max}}d\kappa\,
\frac { 1-\kappa^2(1-ss_1) }{1 -\kappa^2(1-ss_1)^2} \, ,
\end{align}
which then results in 
\begin{align}
    &w_{++}(k,\beta)=w_{--}(k,\beta)= {  \frac{4 L
_x}{\beta }} k''_{\sf max} \, ,
    \nonumber\\
    &w_{+-}(k,\beta)=w_{-+}(k,\beta)={  \frac{2 L
_x}{\beta }}\left(k''_{\sf max}
    -\frac {i\pi  }4k
    \right) .
\end{align}
In both limits the integrals must be real, as they are convergent and the integrand has no imaginary part. This implies that the imaginary part in the result is an artifact of our approximation, and thus we can safely suppress it in what follows. These results can be summarized in the expressions
\begin{equation}
   {  w_{ss_1}(k,\beta) \approx
    \left\{
    \begin{array}{ll}
    \frac{(3+ss_1) }{ 4}
    k''^2_{\sf max}
     & \mbox{low temperatures}
    \\
    ~
    \\
    \frac{(3+ss_1)}{\beta }L_x 
    k''_{\sf max}
     & \mbox{high temperatures}
    \end{array}
    \right.}.
\end{equation}
Notice that $w_{++}(k,\beta)=w_{--}(k,\beta)\equiv W(k,\beta)$ and $w_{-+}(k,\beta)=w_{+-}(k,\beta)=w(k,\beta)
$.
Then the non-vanishing relevant integrals result in
\begin{align}
R_1^{\pm\pm'} &= 2c^2 \left( E_1^{+} \pm (\pm') E_1^{-} \right) 
= -\frac{c^2 L_x^2}{2} \left(  1 \pm (\pm') 1\right) ,
\end{align}
thus
\begin{align}
    R_1^{+-}=R_1^{-+}&=0 \ ,&\quad
    R_1^{--}=R_1^{++}&=-{c^2L_x^2}\equiv R_1 \ .
\end{align}
On the other hand
\begin{align}
R_2^{\pm} &= 4c^2 \left( E_2^{--+} \pm E_2^{++-} \right)  = \frac{ c^2 L_x^4}{16\pi^2} (1 \pm 1)\, f_0(\beta) \ ,
\end{align}
then
\begin{align}
    R_2^-&=0 \ ,\quad&
    R^+_2&=\frac{c^2L_x^4}{8\pi^2}f_0(\beta) \equiv R_2 \  .
\end{align}
Also, we have
\begin{align}
R_3  &=4 p c^2 E_1  =  -p c^2 L_x^2  \ ,
\end{align}
\begin{align}
R_4 &= 2 p c^2 E_1 = -\frac{ p c^2 L_x^2}{2} \ .
\end{align}
Moreover,
\begin{align}
R_5^{\pm\pm'} &= 8c^2 \left[ p^2 E_1 + c^2 \left( E_2^{---} \pm (\pm') E_2^{+++} \right) \right] , \nonumber\\
&= -2 c^2 p^2 L_x^2 
      {-\frac{c^4L_x^3}{ 4\pi}} \left[
 {\frac{L_x}{2\pi}}f_0(\beta)+W(k,\beta) 
\right](1
\pm(\pm')1) \ ,
\end{align}
or, more simply
\begin{align}
R_5^{+-} =R_5^{-+}\equiv R_5
&= -2c^2p^2L_x^2 ,
\nonumber\\
R_5^{++} =R_5^{--}
&= 
 -2 c^2 p^2 L_x^2 
      {-\frac{c^4L_x^3}{ 2\pi}} \left[
 {\frac{L_x}{2\pi}}f_0(\beta)+W(k,\beta) 
\right]
\equiv \tilde R_5 \ .
\end{align}
Finally
\begin{align}
    R_6&=
     -\frac{c^2 p^2 L_x^2 }2
      {-\frac{c^4L_x^3}{ 8\pi}} \left[
 {\frac{L_x}{2\pi}}f_0(\beta)+w(k,\beta) 
\right] .  
\end{align}
With these results, we can now write the $\lambda_{\mu\nu\rho\sigma}$ tensor in the form
\begin{align}
    &\lambda_{zzzz}=g_{zz}g_{zz}'R_6 ,&
    &\lambda_{zzyy}=g_{zz}g_{yy}' R_6 ,\nonumber\\
    &\lambda_{zztx}=-g_{zz}g_{\cdot\cdot}' R_2^-=0 ,&
    &\lambda_{zzty}=-g_{zz} g_{\cdot\theta}'R_4=-g_{zz} g_{\cdot\theta}'R_4 ,\nonumber\\
    &\lambda_{zzxy}=-g_{zz}g_{\cdot y} 'R_4 ,&
    &\lambda_{zzxx}= \lambda_{zztt}=-g_{zz}g_{\cdot\cdot}'R_2^+=-g_{zz}g_{\cdot\cdot}'R_2 , \nonumber\\
    &\lambda_{yyyy} =g_{yy}g_{yy}'R_6 ,&
    &\lambda_{yy tx} =-g_{yy}g_{\cdot\cdot}'R_2^-=0 ,\nonumber\\
    &\lambda_{yy ty} =-g_{yy}g_{\cdot\theta}'R_2^- =0 ,&
    &\lambda_{yyxy}=-g_{yy}g_{\cdot y}'R_2^+ =-g_{yy}g_{\cdot y}'R_2 ,\nonumber\\
    &\lambda_{yyxx} = \lambda_{yy t t} =-g_{yy}g_{\cdot\cdot}'R_2^+=-g_{yy}g_{\cdot\cdot}'R_2 ,&
    &\lambda_{tx tx} =g_{\cdot\cdot}g_{\cdot\cdot}'R_5^{--}=g_{\cdot\cdot}g_{\cdot\cdot}'\tilde R_5 ,\nonumber\\
    &\lambda_{tx ty}=-g_{\cdot\cdot} g_{\cdot y}'R_3^{--}=g_{\cdot\cdot} g_{\cdot y}'R_3 ,&
    &\lambda_{tx xy} =-g_{\cdot\cdot}g_{\cdot y}'R_3^{+-}=-g_{\cdot\cdot}g_{\cdot y}'R_3 ,\nonumber\\
    &\lambda_{tx xx} = \lambda_{tx tt}=g_{\cdot\cdot}g_{\cdot\cdot}'R_5^{-+}=g_{\cdot\cdot}g_{\cdot\cdot}'R_5 ,&
    &\lambda_{ty ty}=g_{\cdot y} g_{\cdot y}'R_1^{--}=g_{\cdot y} g_{\cdot y}'R_1 ,\nonumber\\
    &\lambda_{ty xy}=g_{\cdot y} g_{\cdot y}'R_1^{-+} =0 ,&
    &\lambda_{tyxx}=\lambda_{ty tt}   =g_{\cdot y} g_{\cdot\cdot}'R_3^{-+}=-g_{\cdot y} g_{\cdot\cdot}'R_3 ,\nonumber\\
    &\lambda_{xyxy}=g_{\cdot y} g_{\cdot y}'R_1^{++} =g_{\cdot y} g_{\cdot y}'R_1 ,&
    &\lambda_{xyxx}  =\lambda_{xy tt} =g_{\cdot y} g_{\cdot\cdot}'R_3^{++}=g_{\cdot y} g_{\cdot\cdot}'R_3 ,\nonumber\\
    &\lambda_{xxxx} =\lambda_{xx tt} =\lambda_{tttt} =g_{\cdot\cdot} g_{\cdot\cdot}'R_5^{++}=g_{\cdot\cdot} g_{\cdot\cdot}'\tilde R_5 .
\end{align}
Then, the explicit expressions for the non‑zero components of \(\lambda_{\mu\nu\rho\sigma}\) are
\begin{equation}
\begin{aligned}
\lambda_{zzzz} &= g_{zz}g_{zz}' \left[ -\frac{c^2 p^2 L_x^2}{2} - \frac{c^4 L_x^3}{8\pi}\!\left( \frac{L_x}{2\pi}f_0(\beta) + w(k,\beta) \right) \right], \\[4pt]
\lambda_{zzyy} &= g_{zz}g_{yy}' \left[ -\frac{c^2 p^2 L_x^2}{2} - \frac{c^4 L_x^3}{8\pi}\!\left( \frac{L_x}{2\pi}f_0(\beta) + w(k,\beta) \right) \right], \\[4pt]
\lambda_{zzty}=\lambda_{zzxy} &= \frac{p c^2 L_x^2}{2}\; g_{zz}g_{\cdot y}', \\[4pt]
\lambda_{zzxx} = \lambda_{zztt} &= -\frac{c^2 L_x^4}{8\pi^2}\,f_0(\beta)\; g_{zz}g_{\cdot\cdot}', \\[4pt]
\lambda_{yyyy} &= g_{yy}g_{yy}' \left[ -\frac{c^2 p^2 L_x^2}{2} - \frac{c^4 L_x^3}{8\pi}\!\left( \frac{L_x}{2\pi}f_0(\beta) + w(k,\beta) \right) \right], \\[4pt]
\lambda_{yyxy} &= -\frac{c^2 L_x^4}{8\pi^2}\,f_0(\beta)\; g_{yy}g_{\cdot y}', \\[4pt]
\lambda_{yyxx} = \lambda_{yy tt} &= -\frac{c^2 L_x^4}{8\pi^2}\,f_0(\beta)\; g_{yy}g_{\cdot\cdot}', \\[4pt]
\lambda_{tx ty} &= -p c^2 L_x^2\; g_{\cdot\cdot}g_{\cdot y}', \\[4pt]
\lambda_{tx xy} &= p c^2 L_x^2\; g_{\cdot\cdot}g_{\cdot y}', \\[4pt]
\lambda_{tx xx} = \lambda_{tx tt} &= -2c^2 p^2 L_x^2\; g_{\cdot\cdot}g_{\cdot\cdot}', \\[4pt]
\lambda_{ty ty} =\lambda_{xyxy}&= -c^2 L_x^2\; g_{\cdot y}g_{\cdot y}', \\[4pt]
\lambda_{tyxx} = \lambda_{t y tt} &= p c^2 L_y^2\; g_{\cdot y}g_{\cdot\cdot}', \\[4pt]
\lambda_{xyxx} = \lambda_{xy tt} &= -p c^2 L_x^2\; g_{\cdot y}g_{\cdot\cdot}', \\[4pt]
\lambda_{xxxx} = \lambda_{xx tt} = \lambda_{tx tx} =\lambda_{tttt} &= g_{\cdot\cdot}g_{\cdot\cdot}' \left[ -2c^2 p^2 L_x^2 - \frac{c^4 L_x^3}{2\pi}\!\left( \frac{L_x}{2\pi}f_0(\beta) + W(k,\beta) \right) \right].
\end{aligned}
\end{equation}
From here, we can get the elasticity tensor from the purely spatial components
\begin{equation}
\begin{aligned}
E_{xxxx} &= g_{\cdot\cdot}g_{\cdot\cdot}' \left[ -2c^2 p^2 L_x^2 - \frac{c^4 L_x^3}{2\pi}\!\left( \frac{L_x}{2\pi}f_0(\beta) + W(k,\beta) \right) \right],\\[4pt]
E_{zzzz} &= g_{zz}g_{zz}' \left[ -\frac{c^2 p^2 L_x^2}{2} - \frac{c^4 L_x^3}{8\pi}\!\left( \frac{L_x}{2\pi}f_0(\beta) + w(k,\beta) \right) \right], \\[4pt]
E_{yyyy} &= g_{yy}g_{yy}' \left[ -\frac{c^2 p^2 L_x^2}{2} - \frac{c^4 L_x^3}{8\pi}\!\left( \frac{L_x}{2\pi}f_0(\beta) + w(k,\beta) \right) \right], \\[4pt]
E_{zzyy} &= g_{zz}g_{yy}' \left[ -\frac{c^2 p^2 L_x^2}{2} - \frac{c^4 L_x^3}{8\pi}\!\left( \frac{L_x}{2\pi}f_0(\beta) + w(k,\beta) \right) \right], \\[4pt]
E_{zzxx}  &= -\frac{c^2 L_x^4}{8\pi^2}\,f_0(\beta)\; g_{zz}g_{\cdot\cdot}', \\[4pt]
E_{yyxy} &= -\frac{c^2 L_x^4}{8\pi^2}\,f_0(\beta)\; g_{yy}g_{\cdot y}', \\[4pt]
E_{yyxx}  &= -\frac{c^2 L_x^4}{8\pi^2}\,f_0(\beta)\; g_{yy}g_{\cdot\cdot}', \\[4pt]
E_{xyxx}  &= -p c^2 L_x^2\; g_{\cdot y}g_{\cdot\cdot}'.
\\[4pt]
E_{xyxy}  &= - c^2 L_x^2\; g_{\cdot y}g_{\cdot y}'.
\\[4pt]
E_{xy zz} &= \frac{p c^2 L_x^2}{2}\; g_{\cdot y}g_{zz}'. 
\end{aligned}
\end{equation}
The components of the viscosity tensor vanish. 

For the heat capacity, we obtain
\begin{equation}
     c_V=\frac 1T\,g_{\cdot\cdot}g_{\cdot\cdot}' \left[ 2c^2 p^2 L_x^2 + \frac{c^4 L_x^3}{2\pi}\!\left( \frac{L_x}{2\pi}f_0(\beta) + W(k,\beta) \right) \right] \ , 
\end{equation}
while for the thermal stress tensor we have
\begin{equation}
\begin{aligned}
\beta_{zz} &= -\frac{c^2 L_x^4}{8\pi^2}\,f_0(\beta)\; g_{zz}g_{\cdot\cdot}', \\[4pt]
\beta_{yy} &= -\frac{c^2 L_x^4}{8\pi^2}\,f_0(\beta)\; g_{yy}g_{\cdot\cdot}', \\[4pt]
\beta_{xy} &= -p c^2 L_x^2\; g_{\cdot y}g_{\cdot\cdot}', \\[4pt]
\beta_{xx}    &= g_{\cdot\cdot}g_{\cdot\cdot}' \left[ -2c^2 p^2 L_x^2 - \frac{c^4 L_x^3}{2\pi}\!\left( \frac{L_x}{2\pi}f_0(\beta) + W(k,\beta) \right) \right].
\end{aligned}
\end{equation}
For the thermal inductance, we obtain
\begin{equation}
\begin{aligned}
\iota_{x y} &= \frac 1Tp c^2 L_x^2\; g_{\cdot\cdot}g_{\cdot y}', \\[4pt]
\iota_{yy} &= \frac 1Tc^2 L_x^2\; g_{\cdot y}g_{\cdot y}', \\[4pt]
\iota_{xx}   &=-\frac 1T g_{\cdot\cdot}g_{\cdot\cdot}' \left[ -2c^2 p^2 L_x^2 - \frac{c^4 L_x^3}{2\pi}\!\left( \frac{L_x}{2\pi}f_0(\beta) + W(k,\beta) \right) \right].
\end{aligned}
\end{equation}
The coefficient that gives the heat flux under a homogeneous change in temperature is
\begin{equation}
\begin{aligned}
\pi^{x} &= -\frac 1T2c^2 p^2 L_x^2\; g_{\cdot\cdot}g_{\cdot\cdot}', \\[4pt]
\pi^{y} &= \frac 1Tp c^2 L_x^2\; g_{\cdot y}g_{\cdot\cdot}'. 
\end{aligned}
\end{equation}
Finally, the elastocaloric response is given by
\begin{equation}
\begin{aligned} 
\theta^{zzy} &= \frac{p c^2 L_x^2}{2T}\;g_{\cdot y} g_{zz}', \\[4pt]
\theta^{x xy} &=-\frac 1T p c^2 L_x^2\; g_{\cdot\cdot}g_{\cdot y}', \\[4pt]
\theta^{xxx}  &= \frac 1T2c^2 p^2 L_x^2\; g_{\cdot\cdot}g_{\cdot\cdot}', \\[4pt]
\theta^{yxx} &= -\frac 1Tp c^2 L_x^2\; g_{\cdot y}g_{\cdot\cdot}'. \\[4pt]
\end{aligned}
\end{equation}


\newpage
\begin{thebibliography}{99}

\bibitem{newd3} {\small Y. Nambu and G. Jona-Lasinio, Phys. Rev. 122, 345
(1961); Phys. Rev. 124, 246 (1961). }

\bibitem{newd4} {\small K. Rajagopal and F. Wilczek, in At the Frontier of
Particle Physics/Handbook of QCD, edited by M. Shifman (World Scientific,
Singapore, 2001); arXiv:hep-ph/0011333. }

\bibitem{newd5} {\small M. G. Alford, J. A. Bowers, and K. Rajagopal, Phys.
Rev. D 63, 074016 (2001). }

\bibitem{newd6} {\small R. Casalbuoni and G. Nardulli, Rev. Mod. Phys. 76,
263 (2004). }

\bibitem{pasta1} {\small D.G. Ravenhall, C.J. Pethick, J.R.Wilson, \textit{%
Phys. Rev. Lett.} \textbf{50}, 2066 (1983). }

\bibitem{pasta2} {\small M. Hashimoto, H. Seki, M. Yamada, Prog. Theor.
Phys. 71, 320 (1984). }

\bibitem{pasta2a} {\small C. J. Horowitz, D. K. Berry, C.M. Briggs, M. E.
Caplan, A. Cumming, A. S. Schneider, Phys. Rev. Lett. 114, 031102 (2015). }

\bibitem{pasta2b} {\small D. K. Berry, M. E. Caplan, C. J. Horowitz, G.
Huber, A. S. Schneider, Phys. Rev. C 94, 055801 (2016). }

\bibitem{pasta3} {\small C. O. Dorso, G. A. Frank, J. A. L\'{o}pez, Nucl.
Phys. A978, 35 (2018). }

\bibitem{pasta4} {\small A. da Silva Schneider, M. E. Caplan, D. K. Berry,
C. J. Horowitz, Phys. Rev. C 98, 055801 (2018). }

\bibitem{pasta5} {\small M. E. Caplan, A. S. Schneider, and C. J. Horowitz,
Phys. Rev. Lett. 121, 132701 (2018). }

\bibitem{pasta6} {\small R. Nandi and S. Schramm, J. Astrophys. Astron. 39,
40 (2018). }

\bibitem{pasta7} {\small Z. Lin, M. E. Caplan, C. J. Horowitz, C. Lunardini,
Phys.Rev.C 102 (2020) 4, 045801. }

\bibitem{pasta8} {\small C.O. Dorso, A. Strachan, G.A. Frank, Nucl.Phys.A
1002 (2020) 122004. }

\bibitem{pasta9} {\small C.J. Pethick, Z. Zhang, D.N. Kobyakov, Phys.Rev.C
101 (2020) 5, 055802. }

\bibitem{pasta10} {\small J. A. Lopez, C. O. Dorso, G. A. Frank, Front.Phys.
(Beijing) 16 (2021) 2, 24301. }

\bibitem{aprox0} {\small L. Brey, H. A. Fertig, R. Cote, A. H. MacDonald, 
\textit{Phys. Rev. Lett}. \textbf{75}, 2562 (1995). }

\bibitem{aprox1} {\small I. Klebanov, \textit{Nucl. Phys.} \textbf{B 262}
(1985) 133. }

\bibitem{aprox2} {\small E. Wrist, G.E. Brown, A.D. Jackson, \textit{Nucl.
Phys.} \textbf{A 468} (1987) 450. }

\bibitem{aprox3} {\small N. Manton, \textit{Phys Lett}. \textbf{B 192}
(1987) 177. }

\bibitem{aprox4} {\small A. Goldhaber, N. Manton, \textit{Phys Lett}. 
\textbf{B 198} (1987), 231. }

\bibitem{aprox5} {\small N. Manton, P. Sutcliffe, \textit{Phys. Lett.} 
\textbf{B 342} (1995) 196. }

\bibitem{aprox6} {\small D. Harland, N. Manton, \textit{Nucl. Phys}. \textbf{%
B 935} (2018) 210. }

\bibitem{aprox7} {\small W. K. Baskerville, \textit{Phys. Lett}. \textbf{B
380} (1996) 106. }

\bibitem{aprox8} {\small M. Loewe, C. Villavicencio, Phys. Rev. B 71 (2005)
094001. }

\bibitem{aprox9} {\small M. Loewe, S. Mendizabal, J.C. Rojas, Physics
Letters B 632 (2006) 512--516. }

\bibitem{aprox10} {\small J. A. Ponciano, N. N. Scoccola, Phys. Lett. B 659,
551 (2008). }

\bibitem{pastacond1} {\small D. G. Yakovlev, \textit{Monthly Notices of the
Royal Astronomical Society} \textbf{453}, 581--590 (2015). }

\bibitem{pastacond2} \small{
C.~J.~Horowitz and D.~K.~Berry,
Phys. Rev. C \textbf{78}, 035806 (2008).
}

\bibitem{pastacond3} {\small M.E. Caplan, C.R. Forsman, A.S. Schneider,
Phys.Rev.C 103 (2021) 5, 055810. }

\bibitem{pastacond4} {\small R. Nandi, S. Schramm, J.Astrophys. Astron. 39
(2018) 40; Astrophys.J. 852 (2018) 2, 135. }

\bibitem{pastacond5} {\small H. Sonoda, G. Watanabe, K. Sato, T. Takiwaki,
K. Yasuoka, T. Ebisuzaki, Phys.Rev.C 75 (2007) 042801. }

\bibitem{kubo1} {\small M. S. Green, Melville, \textit{Jour. of Chem. Phys.} 
\textbf{22} (3): 398--413 (1954). }

\bibitem{kubo2} {\small R. Kubo, \textit{Jour. of the Phys. Soc. of Japan}. 
\textbf{12} (6): 570--586 (1957-06-15). }

\bibitem{kubo3} {\small P. C. Martin, J. Schwinger, \textit{Phys. Rev.} 
\textbf{115}, 1342--1373 (1959). }

\bibitem{kubo4} {\small N. Morgenstern Horing, \textit{Quantum Statistical
Field Theory: An Introduction to Schwinger's Variational Method}, Oxford
University Press (2017). }

\bibitem{skyrme} {\small T. H. R. Skyrme, 
Proc. R. Soc. London \textbf{A 260}, 127 (1961). }

\bibitem{Gerard} {\small G. 't Hooft, Nucl. Phy. \textbf{B72}; Nucl. Phys. 
\textbf{B75}, 461 (1974). }

\bibitem{largeN1} {\small G. Veneziano, \textit{Nucl. Phys.} \textbf{B 117},
519 (1976). }

\bibitem{largeN2} {\small E. Witten, \textit{Nucl. Phys.} \textbf{B 160}, 57
(1979). }

\bibitem{CHPT1} {\small S. Scherer, M. R. Schindler, ``\textit{A Primer for Chiral
Perturbation Theory}" Lecture Notes in Physics (Springer-
Verlag, Berlin, Heidelberg, 2012). }

\bibitem{CHPT2} {\small J. Donoghue, E. Golowich, and B. Holstein, ``\textit{Dynamics
of the Standard Model}" (Cambridge University Press,
Cambridge, England, 1994). }

\bibitem{CHPT3} {\small R. Machleidt, D. R. Entem, \textit{Phys. Rep.} \textbf{503}, 1
(2011). }

\bibitem{Lizzi} {\small A.P. Balachandran, A. Barducci, F. Lizzi, C.G.J.
Rodgers, A. Stern, Phys. Rev. Lett. 52 (1984), 887. }

\bibitem{shifman1} {\small M. Shifman, ``\textit{Advanced Topics in Quantum
Field Theory: A Lecture Course}" Cambridge University Press, (2012). }

\bibitem{shifman2} {\small M. Shifman, A. Yung, \textquotedblleft \textit{%
Supersymmetric Solitons}" Cambridge University Press, (2009). }

\bibitem{witten0} {\small E. Witten, 
Nucl. Phys. \textbf{B 223}, 422 (1983); 
Nucl. Phys. \textbf{B 223}, 433 (1983). }

\bibitem{ANW} {\small G. S. Adkins, C. R. Nappi, E. Witten, 
Nucl. Phys. \textbf{B 228}, 552 (1983). }

\bibitem{[8]p} {\small P. Jain, R. Johnson, N.W. Park, J. Schechter, H.
Weigel, Phys. Rev. D 40 (1989) 855. }

\bibitem{[9]p} {\small B. Schwesinger, H. Weigel, G. Holzwarth, A. Hayashi,
Phys. Rep. 173 (1989) 173. }

\bibitem{[10]p} {\small U.-G. MeiBner, Phys. Rep. 161 (1988) 213. }

\bibitem{[11]p} {\small R. Johnson, N.W. Park, J. Schechter, V. Soni, H.
Weigel, Phys. Rev. D 42 (1990) 2998. }

\bibitem{[12]p} {\small D. Masak, Phys. Rev. D 39 (1989) 305. }

\bibitem{sakurai} {\small J. J. Sakurai, \textit{Currents and mesons}.
Chicago: Chicago University Press 1969. }

\bibitem{Machleidt} {\small R.~Machleidt, K.~Holinde and C.~Elster, 
Phys. Rept. \textbf{149} (1987), 1-89. }

\bibitem{m1} {\small G.~S.~Adkins, 
Phys.\ Rev.\ D \textbf{33}, 193 (1986). }

\bibitem{m2} {\small U.~G.~Meissner and I.~Zahed, 
Phys.\ Rev.\ Lett.\ \textbf{56}, 1035 (1986). }

\bibitem{m3} {\small M.~Bando, T.~Kugo, S.~Uehara, K.~Yamawaki and
T.~Yanagida, 
Phys.\ Rev.\ Lett.\ \textbf{54}, 1215 (1985). }

\bibitem{crystal0} {\small P. D. Alvarez, F. Canfora, N. Dimakis, A.
Paliathanasis, \textit{Phys. Lett}. \textbf{B 773}, (2017) 401-407. }

\bibitem{crystal1} {\small F.~Canfora, \textit{Eur.\ Phys.\ J}.\ \textbf{C 78%
}, no. 11, 929 (2018). }

\bibitem{crystal2} {\small F. Canfora, S.-H. Oh, A. Vera, \textit{Eur. Phys. J%
}. \textbf{C 79} (2019) no.6, 485.}

\bibitem{crystal3} {\small F.~Canfora, M.~Lagos and A.~Vera, 
\textit{Eur.\ Phys.\ J.}\ \textbf{C} \textbf{80}, no. 8, 697 (2020). }

\bibitem{crystal4} {\small F.~Canfora, S.~Carignano, M.~Lagos, M.~Mannarelli
and A.~Vera, 
\textit{Phys. Rev.} \textbf{D} 103 (2021) 7, 076003. }

\bibitem{crystal5} {\small P.~D.~Alvarez, S.~L.~Cacciatori, F.~Canfora
and B.~L.~Cerchiai, 
Phys.\ Rev.\ D \textbf{101}, no. 12, 125011 (2020). }

\bibitem{crystal6}
{\small S.~L.~Cacciatori, F.~Canfora, M.~Lagos, F.~Muscolino and A.~Vera,
JHEP \textbf{12}, 150 (2021).}

\bibitem{crystal7}
{\small S.~L.~Cacciatori, F.~Canfora, M.~Lagos, F.~Muscolino and A.~Vera,
Nucl. Phys. B \textbf{976}, 115693 (2022).}

\bibitem{crystal8} {\small F.~Canfora, D.~Hidalgo, M.~Lagos, E.~Meneses
and A.~Vera, 
Phys. Rev. D \textbf{106}, no.10, 105016 (2022). }

\bibitem{crystal9}
F.~Canfora,
JHEP \textbf{11}, 007 (2023).

\bibitem{grav1} E. Ayon-Beato, F. Canfora, J. Zanelli, \textit{Phys.
Lett.} \textbf{B 752, }(2016) 201-205. 

\bibitem{grav2} E.~Ayon-Beato, F.~Canfora, M.~Lagos, J.~Oliva,
A.~Vera, Eur.\ Phys.\ J.\ C \textbf{80}, no. 5, 384 (2020).

\bibitem{Canfora:2025jqr}
F.~Canfora and P.~Pais,
Nucl. Phys. B \textbf{1017}, 116955 (2025).

\bibitem{Canfora:2025qkl}
F.~Canfora and P.~Pais,
Eur. Phys. J. C \textbf{85}, no.8, 884 (2025).

\bibitem{Canfora:2024mkp}
F.~Canfora, M.~Lagos and A.~Vera,
JHEP \textbf{10}, 224 (2024).

\bibitem{Cacciatori:2024mpf}
S.~L.~Cacciatori, F.~Canfora and F.~Muscolino,
Nucl. Phys. B \textbf{1000}, 116477 (2024).

\bibitem{Vera:2025qqz}
A.~Vera,
JHEP \textbf{10}, 119 (2025).

\bibitem{euler4} S.~L.~Cacciatori and A.~Scotti, 
Universe \textbf{8}, no.10, 492 (2022). 

\bibitem{charged1}
L.~Avil{\'e}s, F.~Canfora, N.~Dimakis and D.~Hidalgo,
Phys. Rev. D \textbf{96}, no.12, 125005 (2017).

\bibitem{charged2}
F.~Canfora, M.~Lagos, S.~H.~Oh, J.~Oliva and A.~Vera,
Phys. Rev. D \textbf{98}, no.8, 085003 (2018).

\bibitem{Cacciatori:2025irb}
S.~L.~Cacciatori, F.~Canfora, E.~Delgado, F.~Muscolino and L.~Rosa,
Phys. Rev. D \textbf{113}, no.10, 105025 (2026).

\bibitem{Landau} {\small  L D Landau, E.M. Lifshitz  {\it Statistical Physics,} 
Volume 5 ($3^{rd}$ Edition - January 1, 1980) Paperback ISBN: 9780750633727, eBook ISBN: 9780080570464}

\bibitem{qlo}{\small Podio-Guidugli, P. (2019). 
In: Continuum Thermodynamics. SISSA Springer Series, vol 1. Springer, Cham. https://doi.org/10.1007/978-3-030-11157-1\_5}

\bibitem{thermal} {\small Bosworth, R. C. L. 
Nature 158.4009 (1946): 309-309.}

\bibitem{iesan} {\small
Ieşan, D. 
Z. Angew. Math. Phys. 74, 138 (2023). https://doi.org/10.1007/s00033-023-02034-5}

\bibitem{Kubis} {\small
S.~Kubis and W.~W{\'o}jcik,
Eur. Phys. J. A \textbf{54}, no.12, 215 (2018).}

\end{thebibliography}
\end{document}